\documentclass{conm-p-l}

\usepackage[utf8]{inputenc}

\usepackage{amsmath,amsfonts,amssymb,amsthm}
\usepackage{amsbsy,amsopn,amstext}

\usepackage{tabularx}
\usepackage[charter]{mathdesign}

\usepackage{url}
\usepackage{latexsym}

\usepackage{mathrsfs} 
\usepackage[colorlinks, urlcolor=blue, citecolor=black, linkcolor=black,
breaklinks]{hyperref} 

\usepackage{enumerate}

\usepackage{xypic}
\usepackage{multirow}
\usepackage{array}
\newcolumntype{d}[1]{>{\centering\arraybackslash\hspace{0pt}}m{#1}}

\usepackage{graphicx}

\newcommand{\point}{\mbox{\large\bfseries .}}
 
\newcommand\p{\mathfrak{p}}
 \renewcommand{\cH}{\mathcal{H}}
 \newcommand{\co}{\mathcal{O}}
\newcommand*\ben{\begin{enumerate}} \newcommand\een{\end{enumerate}}
\newcommand*\bit{\begin{itemize}}
\newcommand\eit{\end{itemize}}

\newtheorem{theo}{Theorem}[section]

\newtheorem{ass}[theo]{Assertion}
\newtheorem{lemm}[theo]{Lemma}

\newtheorem{defi}[theo]{Definition}
\newtheorem{propo}[theo]{Proposition}
\newtheorem{coro}[theo]{Corollary}

\newcommand{\Remark}[1][ ]{{\bf Remark. #1 }}

\theoremstyle{remark}

\def\R{\mathbb{R}}
\def\C{\mathbb{C}}
\def\Z{\mathbb{Z}}
\def\Q{\mathbb{Q}}
\def\N{\mathbb{N}}
\def\F{\mathbb{F}}

\def\FF{\mathbf{F}}

\newcommand{\Fq}{\mathbb{F}_q}

\newcommand{\Ld}[2][D]{\ensuremath{\mathscr{L}(#2\mathcal{#1})}} 
\newcommand{\D}[1][D]{\ensuremath{\mathcal{#1}}} 

\newcommand{\mus}{\mu^\mathrm{sym}}

\newcommand{\chch}{{C}hudnovsky-{C}hudnovsky}

\begin{document}

\title{On some bounds for symmetric tensor rank of multiplication in finite
fields}



\author{St\'ephane Ballet}
\address{Aix-Marseille Universit\'e, {CNRS}, Centrale Marseille, Institut de
Math\'ematiques de Marseille \\ Case 907, 163 Avenue de Luminy, F-13288
Marseille Cedex 9, France.}
\email{stephane.ballet@univ-amu.fr}
\thanks{The author wishes to thank the INFRES department for funding his
visit to LTCI, where this work was partly done.}

\author{Julia Pieltant}
\address{{CNRS LTCI}, {T}\'el\'ecom {P}aris{T}ech\\ 46 rue Barrault, F-75634
Paris Cedex 13, France.}
\email{pieltant@enst.fr}
\thanks{This work was supported by a public grant as part of the
Investissement d'avenir project, reference ANR-11-LABX-0056-LMH, LabEx LMH}

\author{Matthieu Rambaud}
\address{{CNRS LTCI}, {T}\'el\'ecom {P}aris{T}ech\\ 46 rue Barrault, F-75634
Paris Cedex 13, France.}
\email{rambaud@enst.fr}

\author{Jeroen Sijsling}
\address{Universit\"at Ulm,
Institut f\"ur Reine Mathematik\\
Helmholtzstrasse 18,
D-89069 Ulm,
Germany.}
\email{sijsling@gmail.com}



\subjclass[2000]{Primary }


\date{\today}

\begin{abstract}
  The aim of this paper is twofold. On the one hand, we establish new upper
  bounds for the symmetric multiplication tensor in any extension of finite
  fields. Note that these bounds are not asymptotic but uniform. On the other
  hand, we clarify the current state of the art by giving the detailed proof of
  some known unpublished uniform bounds, and we discuss the validity of some
  current asymptotic bounds and their relation with the fields of definition of
  certain Shimura curves.
\end{abstract}

\maketitle

\section{Introduction}

\subsection{Tensor rank and symmetric tensor rank}

Let $q$ be a prime power, $\F_q$ be the finite field with $q$ elements and
$\F_{q^n}$ be the degree $n$ extension of $\F_q$. The multiplication of two
elements of $\F_{q^n}$ is a \mbox{$\F_q$-bilinear} application from $\F_{q^n}
\times \F_{q^n}$ onto $\F_{q^n}$. It can\linebreak[4] therefore be considered
as an \mbox{$\F_q$-linear} application from the tensor
product\linebreak[4]${\F_{q^n} \otimes_{\F_q} \F_{q^n}}$ onto $\F_{q^n}$.
Consequently it can be also considered as an element $T$
of\linebreak[4]${{(\F_{q^n} \otimes_{\F_q} \F_{q^n})^\star \otimes_{\F_q}
\F_{q^n}} = {\F_{q^n}^\star \otimes_{\F_q} \F_{q^n}^\star \otimes_{\F_q}
\F_{q^n}}}$. More precisely, when $T$ is written
\begin{equation} \label{tensor}
  T = \sum_{i=1}^{r} x_i^\star\otimes y_i^\star\otimes c_i,
\end{equation}
where the $r$ elements $x_i^\star$ and the $r$ elements $y_i^\star$ are in the
dual $\F_{q^n}^\star$ of $\F_{q^n}$ and the $r$ elements $c_i$ are in
$\F_{q^n}$, the following holds for any ${x,y \in \F_{q^n}}$:
\begin{equation*}
  x \cdot y = \sum_{i=1}^r x_i^\star(x) y_i^\star(y) c_i.
\end{equation*}
The decomposition (\ref{tensor}) is not unique, and neither is the length of
the decomposition \eqref{tensor}. We therefore make the following definition:
\begin{defi}
  The minimal number of summands in a decomposition of the multiplication
  tensor $T$ is called the \emph{bilinear complexity} of the multiplication in
  $\F_{q^n}$ over $\F_q$ and is denoted by $\mu_{q}(n)$:
  \begin{equation*}
    \mu_{q}(n) = \min\left\{r \; \Big| \; T=\sum_{i=1}^{r} x_i^\star\otimes
    y_i^\star\otimes c_i\right\}.
  \end{equation*}
\end{defi}
Hence the bilinear complexity of the multiplication in $\F_{q^n}$ over $\F_q$
is nothing but the rank of the tensor $T$. Among others, a particular class of
decompositions of $T$ is of particular interest, namely the \emph{symmetric
decompositions}:
\begin{equation}\label{symtensor}
  T = \sum_{i=1}^{r} x_i^\star\otimes x_i^\star\otimes c_i.
\end{equation}
 
\begin{defi}
  The minimal number of summands in a symmetric decomposition of the
  multiplication tensor $T$ multiplication is called the \emph{symmetric
  bilinear complexity} of the multiplication in $\F_{q^n}$ over $\F_q$ and is
  denoted by $\mus_{q}(n)$:
  \begin{equation*}
    \mus_{q}(n)= \min\left\{r \; \Big| \; T=\sum_{i=1}^{r} x_i^\star\otimes
    x_i^\star\otimes c_i\right\}.
  \end{equation*}
\end{defi}

One easily gets that ${\mu_q(n) \leq \mus_{q}(n)}$. Some cases where
${\mu_{q}(n) = \mus_{q}(n)}$ are known, but to the best of our knowledge, no
example where ${\mu_{q}(n) < \mus_{q}(n)}$ has been exhibited so far. However,
better upper bounds have been established in the asymmetric case
\cite{rand3,pira} and this may suggest that in general the asymmetric bilinear
complexity of the multiplication and the symmetric one are distinct. In any
case, at the moment, we must consider these two quantities separately.

Note that from an algorithmic point on view as well as for some specific
applications, a symmetric bilinear algorithm can be more interesting than
an\linebreak[4]asymmetric one, unless if {\it a priori}, the constant factor in
the bilinear\linebreak[4] complexity estimation is a little worse. Moreover,
many other research domains are closely related to the determination of
symmetric bilinear multiplication\linebreak[4]algorithms such as, among others,
arithmetic secret sharing and multiparty\linebreak[4]computation (see
\cite{cacrxing3, chcr1})\ldots

\subsection{Known results}\label{sect_known}

The bilinear complexity $\mu_q(n)$ of the multiplication in the $n$-degree
extension of a finite field $\F_q$ is known for certain values of $n$. In
particular, S. Winograd \cite{wino3} and H. de Groote \cite{groo} have shown
that this complexity is ${\geq 2n-1}$, with equality holding if and only if ${n
\leq \frac{1}{2}q + 1}$. Using the principle of the D.V. and G.V. Chudnovsky
algorithm \cite{chch} applied to elliptic curves, M.A. Shokrollahi has shown in
\cite{shok} that the symmetric bilinear complexity of multiplication is equal
to $2n$ for ${\frac{1}{2}q +1< n < \frac{1}{2}(q+1+{\epsilon (q) })}$ where
$\epsilon$ is the function defined by:
\begin{equation*}
  \epsilon (q) = \left \{
    \begin{array}{l}
      \mbox{greatest integer} \leq 2{\sqrt q} \mbox{ prime to $q$, if $q$ is
      not a perfect square} \\
      2{\sqrt q}\mbox{, if $q$ is a perfect square.}
    \end{array}
    \right .
\end{equation*}

Later in \cite{ball1,ball3,balbro,baro1,balb,bach}, the study made by M.A.
Shokrollahi was generalized to algebraic function fields of genus $g$. 

Let us recall that the original algorithm of D.V. and G.V. Chudnovsky
introduced in \cite{chch} is symmetric by definition and leads to the
following result from \cite{ball1}:

\begin{theo}
  Let $q$ be a power of a prime $p$. The symmetric tensor rank $\mus_q(n)$ of
  multiplication in any finite field $\F_{q^n}$ is linear with respect to the
  extension degree; more precisely, there exists a constant $C_q$ such that
  \begin{equation*}
    \mus_q(n) \leq C_q n.
  \end{equation*}
\end{theo}
General expressions for $C_q$ have been obtained, such as the following best
current published estimates:
\begin{equation*}
  C_q=
  \left \{
  \begin{array}{lll}
  \mbox{if } q=2, & \mbox{then} \quad \frac{4824}{247} \simeq 19,6 &
    \mbox{\cite{bapi}  and  \cite{ceoz}} \cr \cr
  \mbox{else if } q=3, &  \mbox{then} \quad 27 & \mbox{\cite{ball3}} \cr \cr
  \mbox{else if } q=p \geq 5, &  \mbox{then} \quad  3\left(1+
    \frac{4}{q-3}\right) & \mbox{\cite{bach}} \cr \cr
  \mbox{else if } q=p^2 \geq 25, & \mbox{then} \quad
    2\left(1+\frac{2}{p-3}\right) & \mbox{\cite{bach}} \cr  \cr
  \mbox{else if } q \geq 4, & \mbox{then} \quad 6\left(1+\frac{p}{q-3}\right) &
    \mbox{\cite{ball3}}
  \end{array}
  \right .
\end{equation*}

Now we introduce the generalized \chch type algorithm described in \cite{ceoz};
the original algorithm given in \cite{chch} by D.V. and G.V. Chudnovsky being
the case where $\deg P_i=1$ and $u_i=1$ for $i=1, \ldots, N$. Here a wider
notion of complexity is involved: the quantity $ \mus_{q}(m, \ell)$, which
corresponds to the symmetric bilinear complexity of the multiplication over
$\Fq$ in $\F_{q^m}[X]/(X^\ell)$, the $\F_q$-algebra of polynomials in one
indeterminate with coefficients in $\F_{q^m}$ truncated at order  $\ell$.

\begin{theo} \label{theo_evalder}
  Let
  \begin{itemize}
    \item $q$ be a prime power,
    \item $\FF/\F_q$ be an algebraic function field,
    \item $Q$ be a degree $n$ place of $F/\F_q$,
    \item ${\mathcal D}$ be a divisor of $F/\F_q$,
    \item ${\mathscr P}=\{P_1,\ldots,P_N\}$ be a set of $N$ places of arbitrary
      degree,
    \item $u_1,\ldots,u_N$ be positive integers.
  \end{itemize}
  We suppose that $Q$ and all the places in $\mathscr P$ are not in the support
  of ${\mathcal D}$ and that:
  \begin{enumerate}[a)]
    \item the map
      \begin{equation*}
        Ev_Q:  \left |
        \begin{array}{ccl}
          \Ld{} & \rightarrow & \F_{q^n}\simeq F_Q \\
          f & \longmapsto & f(Q)
        \end{array}
        \right.
      \end{equation*} 
  is onto,
  \item the map
  \begin{equation*}
    Ev_{\mathscr P} :  \left |
    \begin{array}{ccl}
      \Ld{2} & \longrightarrow & \left(\F_{q^{\deg P_1}}\right)^{u_1} \times
        \left(\F_{q^{\deg P_2}}\right)^{u_2} \times \cdots \times
        \left(\F_{q^{\deg P_N}}\right)^{u_N} \\
      f & \longmapsto & \big(\varphi_1(f), \varphi_2(f), \ldots,
        \varphi_N(f)\big)
    \end{array} \right.
  \end{equation*}
  is injective, where the application $\varphi_i$ is defined by
  \begin{equation*}
    \varphi_i : \left |
    \begin{array}{ccl}
      \Ld{2} & \longrightarrow & \left(\F_{q^{\deg P_i}}\right)^{u_i} \\
      f & \longmapsto & \left(f(P_i), f'(P_i), \ldots, f^{(u_i-1)}(P_i)\right)
    \end{array} \right.
  \end{equation*}
  with $f = f(P_i) + f'(P_i)t_i + f''(P_i)t_i^2+ \ldots + f^{(k)}(P_i)t_i^k +
  \ldots$ the local expansion at $P_i$ of $f$ in ${\Ld{2}}$ with respect to
  the local parameter~$t_i$. Note that we set ${f^{(0)} =f}$.
  \end{enumerate}
  Then 
  \begin{equation*}
    \mus_q(n) \leq \displaystyle \sum_{i=1}^N \mus_q(\deg P_i) \mus_{q^{\deg
      P_i}}(\deg P_i, u_i).
  \end{equation*}
\end{theo}

The following special case of this result was introduced independently by N.
Arnaud in \cite{arna1}, and can be seen as a corollary of Theorem
\ref{theo_evalder} by gathering the places used with the same multiplicity. In
fact ${a_j:= | \{ i\, | \, \deg P_i =j \mbox{ and } u_i=2 \} |}$ for $j = 1,2$
in the statement of the Corollary.

\begin{coro} \label{theo_deg12evalder}
  Let 
  \begin{itemize}
    \item $q$ be a prime power,
    \item $F/\F_q$ be an algebraic function field,
    \item $Q$ be a degree $n$ place of $F/\F_q$, 
    \item $\D$ be a divisor of $F/\F_q$,
    \item ${\mathscr P}=\{P_1,\ldots,P_{N_1}\}$ be a set of $N_1$ places of
      degree one and ${\mathscr P'}=\{R_{1},\ldots,R_{N_2}\}$ be  a set of
      $N_2$ places of degree two,
    \item ${0 \leq a_1 \leq N_1}$ and ${0 \leq a_2 \leq N_2}$ be two integers.
  \end{itemize}
  Suppose that $Q$ and all the places in $\mathscr P$ are not in the support of
  $\D$ and that furthermore
  \begin{enumerate}[a)]
    \item the map
    \begin{equation*}
      Ev_Q: \Ld{} \rightarrow \F_{q^n}\simeq F_Q
    \end{equation*}
    is onto,
    \item the map
    \begin{equation*}
      Ev_{\mathscr P, \mathscr P'}: \left |
      \begin{array}{ccl}
        \Ld{2} & \rightarrow & \F_{q}^{N_1} \times \F_{q}^{a_1}\times
        \F_{q^2}^{N_2} \times \F_{q^2}^{a_2} \\
        f & \mapsto & \big(f(P_1), \ldots, f(P_{N_1}), f'(P_1),\ldots,
          f'(P_{a_1}),\\
        & & \ f(R_{1}), \ldots, f(R_{N_2}), f'(R_{1}),\ldots, f'(R_{a_2})\big)
      \end{array}
      \right .
    \end{equation*}
    is injective.
  \end{enumerate}
  Then  
  \begin{equation*}
    \mus_q(n)\leq  N_1 + 2a_1 + 3N_2 + 6a_2.
  \end{equation*}
\end{coro}

To conclude, we recall some particular exact values for ${\mus_q(n)}$ wich will
be useful for computational use: ${\mu_q(2)=\mus_q(2)=3}$ for any prime power
$q$, ${\mus_2(4)=9}$,\linebreak[4]${\mus_4(4)=\mus_5(4)=8}$ and
${\mus_2(6)=15}$ \cite{chch}.

\subsection{New results and organization of the paper}\label{sect_newresults}

The paper is organized as follows. First we establish new uniform upper bounds
for the tensor rank of multiplication in any finite field, not necessarily of
square cardinality. These bounds are stated in the following theorem:
\begin{theo}\label{theo_arnaudupdate}
  Let ${q=p^r}$ be a power of the prime $p$. Then:
  \begin{enumerate}[(i)]
    \item If ${q\geq 4}$, then $\displaystyle{\mus_{q}(n) \leq 3 \left(1 +
      \frac{\frac{4}{3}p}{(q-3)+2(p-1)\frac{q}{q+1}} \right)n}$.
    \item  If $p\geq5$, then  $\displaystyle{\mus_{p}(n) \leq  3\left(1+
      \frac{8}{3p-5}\right)n}$.
  \end{enumerate}
\end{theo}
These bounds are based on heretofore unpublished work of Arnaud: in fact, we
improve his bounds by using the same general principle, namely the algorithm
that is introduced in Corollary \ref{theo_deg12evalder} applied to two
Garcia-Stichtenoth towers of function fields. Nevertheless, thanks to a more
accurate study of the number of multiplications in the ground field, we are
able to obtain a better bound for $\mus_q(n)$ and $\mus_p(n)$.

Second, we give a detailed proof of two previously known, but also unpublished
bounds that were obtained by Arnaud in his thesis \cite{arna1}. These bounds
hold for extensions of square finite fields and are the following:
\begin{theo}\label{theo_arnaud1}
  Let ${q=p^r}$ be a power of the prime $p$. Then:
  \begin{enumerate}[(i)]
    \item If ${q\geq 4}$, then  $\displaystyle{\mus_{q^2}(n) \leq 2 \left(1 +
      \frac{p}{q-3 + (p-1)\frac{q}{q+1}} \right)n}$.
    \item If $p\geq5$, then  $\displaystyle{\mus_{p^2}(n) \leq 2 \left(1 +
      \frac{2}{p-\frac{33}{16}} \right)n}$.
  \end{enumerate}
\end{theo}
Note that even though bound (i) was established in 2006, it has never been
published in any journal. The proof that is given in this paper is more
complete than the one that can be found in \cite{arna1}. Arnaud also gave
bounds which are similar to bound (ii), but with $p-2$ as denominator.
Unfortunately, this denominator is slightly overestimated under Arnaud's
hypotheses and no calculation is given to prove it in \cite{arna1}. Thus we
give a corrected version with a detailed proof. These two bounds, together with
those of Theorem \ref{theo_arnaudupdate}, rely on a detailed study and careful
calculations in the towers that are presented in \S\ref{sectdeftowers}.

The last section of this paper is devoted to a discussion of an unproven
assumption on a family of Shimura curves that has been used by various authors
to established some asymptotic bounds, admitted to be the current benchmarks.
We first explained how critical the unproven assumption is and give
counter-examples to emphasize its non-triviality. Moreover, we show which
published bounds should no longer be considered as proven.

Our paper therefore consists of two main parts, Section \ref{sect2} and Section
\ref{sect3}, which are widely independent, but both devoted to a reappraisal of
the state of the art of the bounds for the tensor rank in finite fields.

\section{New upper bounds for the symmetric bilinear complexity}\label{sect2}

\subsection{Towers of algebraic function fields}\label{sectdeftowers}

In this section, we introduce some towers of algebraic function fields. An
improved version of Corollary~\ref{theo_deg12evalder} is applied to the
algebraic function fields of these towers to obtain new bounds for the bilinear
complexity. A given curve cannot be used for multiplication in every extension
$\F_{q^n}$ of $\F_q$, but only for $n$ lower than some value. With a tower of
function fields, we can adapt the curve to the degree of the extension. The
important point to note here is that in order to obtain a suitable curve, it
will be desirable to have a tower for which the quotients of two consecutive
genera are as small as possible, or in other words a dense tower. 

For any algebraic function field $F/\F_q$ defined over the finite field $\F_q$,
we denote by $g(F/\F_q)$ the genus of $F/\F_q$ and by $N_k(F/\F_q)$ the number
of places of degree $k$ in $F/\F_q$.

\subsubsection{A Garcia-Stichtenoth tower of Artin-Schreier function field
extensions}

We now present a modified Garcia-Stichtenoth tower (cf.
\cite{gast,ball3,baro1}) having good properties. Let us consider a finite field
$\F_{q^2}$ with $q = p^r>3$ and let $T_1$ be the elementary abelian tower over
$\F_{q^2}$ after Garcia-Stichtenoth \cite{gast}. This is defined by the
sequence $(F_1, F_2,\ldots)$ where 
\begin{equation*}
  F_{k+1} := F_{k}(z_{k+1})
\end{equation*} 
and $z_{k+1}$ satisfies the equation: 
\begin{equation*}
  z_{k+1}^q+z_{k+1}=x_k^{q+1}
\end{equation*} 
with 
\begin{equation*}
  x_k := z_k/x_{k-1}\mbox{ in } F_k \mbox{ (for $k\geq2$).}
\end{equation*}
Moreover, $F_1:=\F_{q^2}(x_1)$ is the rational function field over $\F_{q^2}$
and $F_2$ the Hermitian function field over $\F_{q^2}$. Let us denote by $g_k$
the genus of $F_k$. Then we have the following formulae:
\begin{equation}\label{genregs}
g_k = \left\{
  \begin{array}{ll}
    q^k+q^{k-1}-q^\frac{k+1}{2} - 2q^\frac{k-1}{2}+1 & \mbox{if } k \equiv 1
      \mod 2,\\
    q^k+q^{k-1}-\frac{1}{2}q^{\frac{k}{2}+1} -
      \frac{3}{2}q^{\frac{k}{2}}-q^{\frac{k}{2}-1} +1& \mbox{if } k \equiv 0
      \mod 2.
  \end{array}
  \right .
\end{equation}
As in \cite{ball3}, let us consider the completed Garcia-Stichtenoth tower 
\begin{equation*}
  T_2 = F_{1,0}\subseteq F_{1,1}\subseteq \cdots \subseteq F_{1,r}
      = F_{2,0}\subseteq F_{2,1} \subseteq \cdots \subseteq F_{2,r} \subseteq
        \cdots .
\end{equation*} 
It has the property that $F_k \subseteq F_{k,s} \subseteq F_{k+1}$ for any
integer $s \in \{0,\ldots,r\}$,  where $F_{k,0}=F_k$ and $F_{k,r}=F_{k+1}$.
Recall that each extension $F_{k,s}/F_k$ is Galois of degree $p^s$ with full
constant field $\F_{q^2}$. Now consider the tower studied in \cite{baro1}:
\begin{equation*}
  T_3 = G_{1,0} \subseteq G_{1,1} \subseteq \cdots \subseteq G_{1,r}=
  G_{2,0}\subseteq G_{2,1}\subseteq \cdots \subseteq G_{2,r} \subseteq \cdots
\end{equation*}
defined over the constant field $\F_q$. It is related to  the tower $T_2$ by
\begin{equation*}
  F_{k,s} = \F_{q^2}G_{k,s} \quad \mbox{for all $k$ and $s$.}
\end{equation*}
In other words, $F_{k,s}$ can be obtained from $G_{k,s}$ by extending the
constant field from $\F_q$ to $\F_{q^2}$. Note that the tower $T_3$ is
well-defined by \cite{baro1} and \cite{balbro}. Moreover, we have the following
result:

\begin{propo}\label{subfield}
  Let ${q = p^r \geq4}$ be a prime power. For all integers $k \geq 1$ and ${s
  \in \{0, \ldots,r\}}$, there exists a step $F_{k,s}/\F_{q^2}$ (respectively
  $G_{k,s}/\F_q$) with genus $g_{k,s}$ and $N_{k,s}$ rational places in
  $F_{k,s}/\F_{q^2}$ (respectively ${N_{k,s} = N_1(G_{k,s}/\F_q) + 2 N_2
  (G_{k,s}/\F_q)}$) such that:
\begin{enumerate}[(1)]
  \item $F_k \subseteq F_{k,s} \subseteq F_{k+1}$, where we set $F_{k,0}=F_k$
    and $F_{k,r}=F_{k+1}$,\\(respectively $G_k \subseteq G_{k,s} \subseteq
    G_{k+1}$, where we set $G_{k,0}=G_k$ and $G_{k,r}=G_{k+1}$),
  \item $\big( g_k-1 \big)p^s +1 \leq g_{k,s} \leq \frac{g_{k+1}}{p^{r-s}} +1$,
  \item $N_{k,s} \geq (q^2-1)q^{k-1}p^s$.
\end{enumerate} 
\end{propo}

\subsubsection{A Garcia-Stichtenoth tower of Kummer function field extensions}

In this section, we present a Garcia-Stichtenoth tower (cf. \cite{bach}) having
good properties. Let $\F_q$ be a finite field of characteristic $p\geq3$. Let
us consider the tower $T$ over $\F_q$ which is defined recursively by the
following equation, studied in \cite{gast2}:
\begin{equation*}
  y^2=\frac{x^2+1}{2x}.
\end{equation*}
The tower $T/\F_q$ is represented by the sequence of function fields $(H_0,
H_1, H_2,\ldots)$ where $H_n = \F_q(x_0, x_1,\ldots, x_n)$ and
$x_{i+1}^2=(x_i^2+1)/2x_i$ holds for each $i\geq 0$. Note that $H_0$ is the
rational function field. For any prime number $p \geq 3$, the tower
$T/\F_{p^2}$ is asymptotically optimal over the field $\F_{p^2}$, i.e.
$T/\F_{p^2}$ reaches the Drinfeld-Vladut bound. Moreover, for any integer $k$,
$H_k/\F_{p^2}$ is the constant field extension of $H_k/\F_p$.

From \cite{bach}, we know that the genus $g(H_k)$ of the step $H_k$ is given by:
\begin{equation}\label{genregsr}
  g(H_k) = \left\{
    \begin{array}{ll}
      2^{k+1}-3\cdot 2^\frac{k}{2}+1 & \mbox{if } k \equiv 0 \mod 2,\\
      2^{k+1} -2\cdot 2^\frac{k+1}{2}+1& \mbox{if } k \equiv 1 \mod 2.
    \end{array}
  \right .
\end{equation}
and that the following bounds hold for the number of rational places  in $H_k$
over $\F_{p^2}$ and for  the number of places of degree 1 and 2 over $\F_p$:
\begin{equation}\label{nbratplgsr}
  N_1(H_k/\F_{p^2}) \geq 2^{k+1}(p-1)
\end{equation}
and
\begin{equation}\label{nbpldeg12gsr}
  N_1(H_k/\F_p) +2N_2(H_k/\F_p) \geq 2^{k+1}(p-1).
\end{equation}

\subsection{Some preliminary results} \label{sectionusefull}

We now proceed to establish some technical results on the genus and number of
places of the steps of the towers $T_2/\F_{q^2}$, $T_3/\F_q$, $T/\F_{p^2}$ and
$T/\F_p$ defined in the previous section. These results will allow us to
determine a suitable step of the tower to which we can apply the algorithm.

\subsubsection{About the Garcia-Stichtenoth tower of Artin-Schreier extensions}

In this section, $q:=p^r$ is a power of the prime $p$. 

\begin{lemm}\label{lemme_genre}
  Let ${q>3}$. We have the following bounds on the genus for the steps of the
  towers $T_2/\F_{q^2}$ and  $T_3/\F_q$:
  \begin{enumerate}[i)]
    \item $g_k> q^k$ for all ${k\geq 4}$,
    \item $g_k \leq q^{k-1}(q+1) - \sqrt{q}q^\frac{k}{2}$,
    \item $g_{k,s} \leq q^{k-1}(q+1)p^s$ for all ${k\geq 0}$ and
      $s=0,\ldots,r$,
    \item $g_{k,s} \leq \frac{q^k(q+1)-q^\frac{k}{2}(q-1)}{p^{r-s}}$ for all
      $k\geq 2$ and $s=0,\ldots,r$.
  \end{enumerate}
\end{lemm}

\begin{Proof}
  \textit{(i)} According to Formula (\ref{genregs}), we know that if $k$ is odd,
  then 
  \begin{equation*}
    g_k = q^k+q^{k-1}-q^\frac{k+1}{2} - 2q^\frac{k-1}{2} + 1
        = q^k+q^\frac{k-1}{2}(q^\frac{k-1}{2} - q - 2) + 1.
  \end{equation*}
  Since ${q>3}$ and ${k \geq 4}$, we have ${q^\frac{k-1}{2} - q - 2 >0}$, thus
  ${g_k>q^k}$.

  On the other hand, if $k$ is even, then 
  \begin{equation*}
    g_k = q^k+q^{k-1}-\frac{1}{2}q^{\frac{k}{2}+1} -
            \frac{3}{2}q^{\frac{k}{2}}-q^{\frac{k}{2}-1} +1
        = q^k+q^{\frac{k}{2}-1}(q^\frac{k}{2}-\frac{1}{2}q^{2} -
            \frac{3}{2}q-1)+1.
  \end{equation*}
  Since ${q>3}$ and ${k\geq 4}$, we have ${q^\frac{k}{2}-\frac{1}{2}q^{2} -
  \frac{3}{2}q-1>0}$, thus ${g_k>q^k}$.

  \textit{(ii)} This follows from Formula (\ref{genregs}) since for all $k\geq
  1$ we have ${2q^\frac{k-1}{2} \geq 1}$, which deals with the case of odd $k$,
  and ${\frac{3}{2}q^\frac{k}{2}+q^{\frac{k}{2}-1}\geq 1}$ which deal with the
  case of even $k$ since ${\frac{1}{2}q\geq \sqrt{q}}$.

  \textit{(iii)} If ${s=r}$, then according to Formula (\ref{genregs}), we have 
  \begin{equation*}
    g_{k,s} = g_{k+1}\leq q^{k+1}+q^{k} = q^{k-1}(q+1)p^s.
  \end{equation*}
  Otherwise we have that ${s<r}$. Then Proposition \ref{subfield} says that
  ${g_{k,s} \leq \frac{g_{k+1}}{p^{r-s}}+1}$. Moreover, since
  ${q^\frac{k+2}{2}\geq q}$ and ${\frac{1}{2}q^{\frac{k+1}{2}+1}\geq q}$, we
  obtain ${g_{k+1}\leq q^{k+1} + q^k - q + 1}$ from Formula (\ref{genregs}).
  Thus we get 
  \begin{eqnarray*}
    g_{k,s} & \leq & \frac{q^{k+1} + q^k - q + 1}{p^{r-s}} +1\\
    &  = & q^{k-1}(q+1)p^s - p^s + p^{s-r} + 1\\
    & \leq & q^{k-1}(q+1)p^s + p^{s-r}\\
    & \leq & q^{k-1}(q+1)p^s \  \mbox{ since ${0 \leq p^{s-r} <1}$ and
      ${g_{k,s} \in \N}$}.
  \end{eqnarray*}

  \textit{(iv)} This follows from ii) since Proposition \ref{subfield} gives
  ${g_{k,s} \leq \frac{g_{k+1}}{p^{r-s}}+1}$, so \linebreak[4]${g_{k,s} \leq
  \frac{q^{k}(q+1) - \sqrt{q}q^\frac{k+1}{2}}{p^{r-s}} +1}$ which gives the
  result since ${p^{r-s} \leq q^\frac{k}{2}}$ for all ${k\geq2}$.
  \qed
\end{Proof}

\begin{lemm}{\label{lemme_delta}}
  Let $q>3$ and $k\geq4$. We set ${\Delta g_{k,s} := g_{k,s+1} - g_{k,s}}$,
  ${D_{k,s}:=(p-1)p^sq^k}$ and ${N_{k,s} := N_1(F_{k,s}/\F_{q^2}) =
  N_1(G_{k,s}/\F_q)+2N_2(G_{k,s}/\F_q)}$. One has:
  \begin{enumerate}[(i)]
    \item $\Delta g_{k,s} \geq D_{k,s}$,
    \item $N_{k,s} \geq D_{k,s}$.
  \end{enumerate}
\end{lemm}

\begin{Proof}
  \textit{(i)} From the Riemann--Hurwitz Formula, one has ${g_{k,s+1}-1 \geq
  p(g_{k,s}-1)}$, so ${g_{k,s+1}-g_{k,s} \geq (p-1)(g_{k,s}-1)}$. Applying the
  Riemann--Hurwitz formula $s$ more times, we get ${g_{k,s+1}-g_{k,s} \geq
  (p-1)p^s\big(g(G_k)-1\big)}$. Thus Lemma~\ref{lemme_genre}(i) gives that\linebreak[4]
  ${g_{k,s+1}-g_{k,s} \geq (p-1)p^sq^k}$ since $q>3$ and $k\geq 4$.

  \textit{(ii)} According to Proposition \ref{subfield}, one has
  \begin{eqnarray*}
    N_{k,s} & \geq & (q^2-1)q^{k-1}p^s \\
    & = & (q+1)(q-1)q^{k-1}p^s \\
    & \geq & (q-1)q^kp^s\\
    & \geq & (p-1)q^kp^s \mbox{.}
  \end{eqnarray*}
  \qed
\end{Proof}

\begin{lemm}\label{lemme_bornesup}
  Let ${N_{k,s} := N_1(F_{k,s}/\F_{q^2}) =
  N_1(G_{k,s}/\F_q)+2N_2(G_{k,s}/\F_q)}$. For all ${k \geq 1}$ and ${s=0,
  \ldots, r}$, we have
  \begin{equation*}
    \sup \big \{ n \in \N \; | \;  2n \leq N_{k,s} -2g_{k,s} +1 \big \} \geq
    \frac{1}{2}(q+1)q^{k-1}p^s(q-3).
  \end{equation*}
\end{lemm}

\begin{Proof}
  From Proposition \ref{subfield} and Lemma \ref{lemme_genre} iii), we get 
  \begin{eqnarray*}
    N_{k,s} -2g_{k,s} +1 & \geq & (q^2-1)q^{k-1}p^s - 2q^{k-1}(q+1)p^s +1 \\
    & = & (q+1)q^{k-1}p^s\big((q-1) -2\big) +1 \\
    & \geq & (q+1)q^{k-1}p^s(q-3)
  \end{eqnarray*}
  thus we have $\sup \big \{ n \in \N \; | \;  2n \leq N_{k,s} -2g_{k,s} +1
  \big \} \geq \frac{1}{2}q^{k-1}p^s(q+1)(q-3)$.
  \qed
\end{Proof}

\subsubsection{About the Garcia-Stichtenoth tower of Kummer extensions}
In this section, $p$ is an odd prime. We denote by $g_k$ the genus of the step
$H_k$ and we fix \linebreak[4]${N_k := N_1(H_k/\F_{p^2}) = N_1(H_k/\F_p) +
2N_2(H_k/\F_p)}$. The following lemma follows from Formulae~(\ref{genregsr})
and (\ref{nbpldeg12gsr}):
\begin{lemm}\label{lemme_genregsr}
  These two bounds hold for the genus of each step of the towers $T/\F_{p^2}$
  and $T/\F_p$:
  \begin{enumerate}[i)]
    \item $g_k \leq 2^{k+1}-2\cdot2^\frac{k+1}{2}+1$,
    \item $g_k \leq 2^{k+1}$.
  \end{enumerate}
\end{lemm}

\begin{lemm}\label{lemme_deltagsr}
  For all $k\geq 0$, we set ${\Delta g_k := g_{k+1} - g_k}$. Then one has
  \linebreak[4]${\Delta g_k \geq 2^{k+1}- 2^\frac{k+1}{2}}$ and ${N_k\geq
  \frac{4}{3}\Delta g_k}$ (so we also have that ${N_k\geq \Delta g_k}$).
\end{lemm}

\begin{Proof}
  If $k$ is even then ${\Delta g_k = 2^{k+1}-2^\frac{k}{2}}$, else ${\Delta g_k
  = 2^{k+1}-2^\frac{k+1}{2}}$ so the first equality holds trivially.
  Moreover, since ${p\geq 3}$, the first second follows from the bounds
  (\ref{nbratplgsr}) and (\ref{nbpldeg12gsr}) which gives ${N_k \geq 2^{k+2}>
  2\Delta g_k}$. \qed
\end{Proof}

\begin{lemm}\label{lemme_bornesupgsr}
  Let $H_k$ be a step of one of the towers $T/\F_{p^2}$ or $T/\F_p$. One has:
  \begin{equation*}
  \sup \big \{ n \in \N \; | \;  N_k \geq 2n +2g_k -1\big \} \geq 2^{k}(p-3)+2.
  \end{equation*}
\end{lemm}

\begin{Proof}
  From the bounds (\ref{nbratplgsr}) and (\ref{nbpldeg12gsr}) for $N_k$ and
  Lemma \ref{lemme_genregsr}(i), we get 
  \begin{eqnarray*}
    N_k - 2g_k +1 & \geq & 2^{k+1}(p-1) -2(2^{k+1}-2\cdot2^\frac{k+1}{2}+1)+1\\
    & = & 2^{k+1}(p-3) + 4\cdot2^\frac{k+1}{2} - 1 \\
    & \geq & 2^{k+1}(p-3) + 4 \mbox{ since } k\geq0.
  \end{eqnarray*}\qed
\end{Proof}

\subsection{General results for $\mus_q(n)$}
In \cite{balb}, Ballet and Le Brigand proved the following useful result:
\begin{theo}\label{existdivnonspe}
  Let $F/\F_q$ be an algebraic function field of genus $g\geq 2$. If $q\geq4$,
  then there exists a non-special divisor of degree $g-1$.
\end{theo}

The four following lemmas prove the existence of a ``good'' step of the towers
defined in \S\ref{sectdeftowers}, that is to say a step that will be optimal
for the bilinear complexity of multiplication:

\begin{lemm}\label{lemme_placedegn}
  Let $n \geq \frac{1}{2}\left(q^2+1+\epsilon(q^2)\right)$ be an integer. If
  $q=p^r\geq4$, then there exists a step $F_{k,s}/\F_{q^2}$ of the tower
  $T_2/\F_{q^2}$ such that all the three following conditions are verified:
  \begin{enumerate}[(1)]
    \item there exists a non-special divisor of degree $g_{k,s}-1$ in
      $F_{k,s}/\F_{q^2}$,
    \item there exists a place of $F_{k,s}/\F_{q^2}$ of degree $n$,
    \item $N_1(F_{k,s}/\F_{q^2}) \geq 2n + 2g_{k,s}-1$.
  \end{enumerate}
  Moreover, the first step for which both Conditions (2) and (3) are verified
  is the first step for which (3) is verified.
\end{lemm}

\begin{Proof}
  Note that $n \geq 9$ since $q\geq4$ and ${n \geq \frac{1}{2}(q^2+1) \geq
8.5}$. Fix $1 \leq k \leq n-4$ and ${s \in \{0, \ldots, r\}}$. First, we prove
that condition (2) is verified. Lemma~\ref{lemme_genre}(iv) gives:
  \begin{eqnarray}
    \nonumber 2g_{k,s}+1 & \leq & 2\frac{q^k(q+1) -
      q^\frac{k}{2}(q-1)}{p^{r-s}} + 1\\
    \nonumber & = & 2p^s\left(q^{k-1}(q+1)-q^\frac{k}{2}\frac{q-1}{q}\right)
      +1\\
    & \leq & 2q^{k-1}p^s(q+1) \ \  \ \mbox{\ since }
      2p^sq^\frac{k}{2}\frac{q-1}{q}\geq1 \label{eqgenre1}\\
    \nonumber & \leq & 2q^k(q^2-1).
  \end{eqnarray}
  On the other hand, one has ${n-1 \geq k+3 > k+\frac{1}{2}+2}$ so $n-1 \geq
  \log_q(q^k)+\log_q(2)+\log_q(q+1)$. This gives ${q^{n-1} \geq 2q^k(q+1)}$,
  hence $q^{n-1}(q-1) \geq 2q^k(q^2-1)$. Therefore, one has ${2g_{k,s}+1 \leq
  q^{n-1}(q-1)}$, which ensures that condition (2) is satisfied according to
  Corollary 5.2.10 in \cite{stic}.

  Now suppose in addition that ${k \geq \log_q\left(\frac{2n}{5}\right)+1}$.
  Note that for all $n\geq 9$ there exists such an integer $k$ since the size
  of the interval $[\log_q\left(\frac{2n}{5}\right)+1 , n-4]$ is bigger
  than ${9-4-\log_4\left(\frac{2\cdot9}{5}\right)-1 \geq 3 >1}$.
  Moreover, such an integer $k$ verifies ${q^{k-1} \geq \frac{2}{5}n}$, so ${n
  \leq \frac{1}{2}q^{k-1}(q+1)(q-3)}$ since $q\geq4$. Then one has
  \begin {eqnarray*}
    2n+2g_{k,s}-1 & \leq & 2n+2g_{k,s}+1\\
    & \leq & 2n + 2q^{k-1}p^s(q+1) \ \  \ \mbox{\ according to
      (\ref{eqgenre1})}\\
    & \leq & q^{k-1}(q+1)(q-3) + 2q^{k-1}p^s(q+1)\\
    & \leq  & q^{k-1}p^s(q+1)(q-1) \\
    & = & (q^2-1)q^{k-1}p^s
  \end{eqnarray*}
  which gives ${N_1(F_{k,s}/\F_{q^2}) \geq 2n + 2g_{k,s}-1}$ according to
  Proposition \ref{subfield} (3). Hence, for any integer $k \in
  [\log_q\left(\frac{2n}{5}\right)+1 , n-4]$, conditions (2) and (3) are
  satisfied and the smallest integer $k$ for which they are both satisfied is
  the smallest integer $k$ for which condition (3) is satisfied
  
  To conclude, remark that for such an integer $k$, condition (1) is easily
  verified by using Theorem \ref{existdivnonspe}, since $q\geq 4$ and ${g_{k,s}
  \geq g_2\geq 6}$ according to Formula (\ref{genregs}). \\  \qed
\end{Proof} 
 
The similar result for the tower $T_3/\F_q$ is as follows:
\begin{lemm}\label{lemme_placedegn2}
  Let $n \geq \frac{1}{2}\left(q+1+\epsilon(q)\right)$ be an integer. If
  $q=p^r\geq 4$, then there exists a step $G_{k,s}/\F_q$ of the tower
  $T_3/\F_q$ such that all the three following conditions are verified:
  \begin{enumerate}[(1)]
    \item there exists a non-special divisor of degree $g_{k,s}-1$ in
      $G_{k,s}/\F_q$,
    \item there exists a place of $G_{k,s}/\F_q$ of degree $n$, 
    \item $N_1(G_{k,s}/\F_q)+2N_2(G_{k,s}/\F_q) \geq 2n + 2g_{k,s}-1$.
  \end{enumerate}
  Moreover, the first step for which both Conditions (2) and (3) are verified
  is the first step for which (3) is verified.
\end{lemm}

\begin{Proof}
Note that $n \geq 5$ since $q\geq4$, ${\epsilon(q) \geq \epsilon(4)=4}$ and ${n
\geq \frac{1}{2}(q+1+\epsilon(q)) \geq 4.5}$. First we focus on the case
$n\geq12$. Fix $1 \leq k \leq \frac{n-5}{2}$ and ${s \in \{0, \ldots, r\}}$.
One has ${2p^sq^k\frac{q+1}{\frac{\sqrt{q}}{2}} \leq q^\frac{n-1}{2}}$ since
\begin{equation*}{
  \frac{n-1}{2} \geq k + 2 = k -\frac{1}{3} +1+1+\frac{3}{2} \geq
  \log_q(q^{k-\frac{3}{2}}) +\log_q(4)+\log_q(p^s)+\log_q(q+1)}.
\end{equation*}
Hence ${2p^sq^{k-1}(q+1) \leq q^\frac{n-1}{2}(\sqrt{q}-1)}$ since
${\frac{\sqrt{q}}{2}\leq \sqrt{q}-1}$ for $q\geq4$. According to
(\ref{eqgenre1}) in the previous proof, this proves that condition (2) is
satisfied.

The same reasoning as in the previous proof shows that condition (3) is also
satisfied as soon as ${k \geq \log_q\left(\frac{2n}{5}\right)+1}$. Moreover,
for $n\geq12$, the interval\linebreak[4]${[\log_q\left(\frac{2n}{5}\right)+1 ,
\frac{n-7}{2}]}$ contains at least one integer, and the smallest integer $k$ in
this interval is the smallest integer $k$ for which condition (3) is verified.
Furthermore, for such an integer $k$, condition (1) is easily verified from
Theorem \ref{existdivnonspe} since $q\geq 4$ and ${g_{k,s} \geq g_2\geq 6}$
according to Formula (\ref{genregs}).

To complete the proof, we deal with case $5\leq n\leq11$. For this
case, we have to look at the values of $q=p^r$ and $n$ for which we have both
${n\geq \frac{1}{2}\left(q+1+\epsilon(q)\right)}$ and ${5 \leq n \leq 11}$. For
each value of $n$ such that these two inequalities are satisfied, we have to
check that conditions (1), (2) and (3) are verified. In this aim, we use the
KASH packages \cite{kash} to compute the genus and number of places of degree 1
and 2 of the first steps of the tower $T_3/\F_q$. Thus we determine the first
step $G_{k,s}/\F_q$ that satisfied all the three conditions (1), (2) and (3).
We resume our results in the following table: \hspace{-8em} 

\begin{equation*}
  \begin{array}{|c|c|c|c|}
    \hline
    q=p^r & 2^2 & 2^3 & 3^2 \\
    \hline
    \epsilon(q) & 4 & 5 & 6 \\
    \hline
    \frac{1}{2}\left(q+1+\epsilon(q)\right) & 4.5 & 7 & 8  \\
    \hline
    n \mbox{ to be considered} & 5 \leq n \leq 11 & 7 \leq n \leq 11 & 8 \leq n
      \leq 11 \\
    \hline
    (k,s) & (1,1) & (1,1) & (1,1) \\
    \hline
    N_1(G_{k,s}/\F_q) & 5 & 9 & 10  \\
    \hline
    N_2(G_{k,s}/\F_q) & 14 & 124 & 117 \\
    \hline
    \Gamma(G_{k,s}/\F_q) & 15 & 117 & 113 \\
    \hline
    g_{k,s} & 2 & 12 & 9 \\
    \hline
    2g_{k,s}+1 & 5 & 25 & 19 \\
    \hline
    q^\frac{n-1}{2}(\sqrt{q}-1) \geq \ldots & 16 & 936 & 4374 \\
    \hline
  \end{array}
\end{equation*}

\vspace{-0.5em}
\hspace{-8em} 
\begin{equation*}
  \begin{array}{|c|c|c|c|c|}
    \hline
    q=p^r & 5 & 7 & 11 & 13 \\
     \hline
    \epsilon(q) & 4 & 5 & 6 & 7\\
    \hline
    \frac{1}{2}\left(q+1+\epsilon(q)\right) & 5 & 6.5 & 9 & 10.5 \\
    \hline
    n \mbox{ to be considered} &  5 \leq n \leq 11 & 7 \leq n \leq 11 & 9 \leq
      n \leq 12 &  n = 11 \\
    \hline
    (k,s) & (2,0) & (2,0) & (2,0) & (2,0) \\
    \hline
    N_1(G_{k,s}/\F_q) & 6 & 8 & 12 & 14 \\
    \hline
    N_2(G_{k,s}/\F_q) & 60 & 168 & 660 & 1092 \\
    \hline
    \Gamma(G_{k,s}/\F_q) & 53 & 151.5  & 611.5  & 1021.5  \\
    \hline
    g_{k,s} & 10 & 21 & 55 & 78 \\
    \hline
    2g_{k,s}+1 & 21 & 43 & 11 & 157 \\
    \hline
    q^\frac{n-1}{2}(\sqrt{q}-1) \geq \ldots & 30 & 564 & 33917 & 967422 \\
    \hline
  \end{array}
\end{equation*}
\vspace{1em}

In this table, one can check that for each value of $q$ and $n$ to be
considered and every corresponding step $G_{k,s}/\F_q$ one has simultaneously:
\begin{itemize}
  \item $g_{k,s}\geq2$ so condition (1) is verified according to Theorem
    \ref{existdivnonspe},
  \item $2g_{k,s}+1 \leq q^\frac{n-1}{2}(\sqrt{q}-1)$ so condition (2) is
    verified.
  \item $\Gamma(G_{k,s}/\F_q) := \frac{1}{2} \left(N_1(G_{k,s}/\F_q) + 2N_2
    (G_{k,s}/\F_q) - 2g_{k,s} + 1 \right) \geq n$, so condition (3) is
    verified.
\end{itemize}
\qed
\end{Proof}

The similar result for the tower $T/\F_{p^2}$ is as follows:

\begin{lemm}\label{lemme_placedegngsr}
  Let $p\geq5$ and $n \geq \frac{1}{2}\left(p^2+1+\epsilon(p^2)\right)$. There
  exists a step $H_k/\F_{p^2}$ of the tower $T/\F_{p^2}$ such that the three
  following conditions are verified:
  \begin{enumerate}[(1)]
    \item there exists a non-special divisor of degree $g_k-1$ in
      $H_k/\F_{p^2}$,
    \item there exists a place of $H_k/\F_{p^2}$ of degree $n$,
    \item $N_1(H_k/\F_{p^2}) \geq 2n + 2g_k - 1$.
  \end{enumerate}
  Moreover, the first step for which all the three conditions are verified is
  the first step for which (3) is verified.
\end{lemm}

\begin{Proof}
  Note that ${n \geq \frac{1}{2}(5^2+1+\epsilon(5^2)) = 18}$. We first prove
  that for all integers $k$ such that ${2 \leq k \leq n - 2}$, we have
  ${2g_k+1 \leq p^{n-1}(p-1)}$ , so condition (2) is verified according to
  Corollary 5.2.10 in \cite{stic2}. Indeed, for such an integer $k$,
  since\linebreak[4] ${p\geq5}$ one has ${k \leq \log_2(p^{n-2}) \leq
  \log_2(p^{n-1}-1)}$, thus it holds that \linebreak[4]${k+2 \leq
  \log_2\left(4(p^{n-1}-1)\right) \leq \log_2 (4p^{n-1}-1)}$ and then
  ${2^{k+2}+1 \leq 4p^{n-1}}$. Hence\linebreak[4]${2\cdot2^{k+1} +1 \leq
  p^{n-1}(p-1)}$ since ${p\geq5}$, which gives the result according to
  \linebreak[4]Lemma~\ref{lemme_genregsr}(ii).

  We prove now that for ${k\geq \log_2 (2n-1)-2}$, condition (3) is verified.
  Indeed, for such an integer $k$, we have ${k +2\geq \log_2 (2n-1)}$, so
  ${2^{k+2} \geq 2n-1}$. Hence we get ${2^{k+3} \geq 2n+2^{k+2}-1}$ and so
  ${2^{k+1}(p-1)\geq 2^{k+1}\cdot4 \geq 2n+2^{k+2}-1}$ since ${p\geq5}$. Thus
  we have ${N_1(H_k/\F_{p^2}) \geq 2n + 2g_k - 1}$ according to the bound
  (\ref{nbratplgsr}) and\linebreak[4]Lemma~\ref{lemme_genregsr}(ii).

  Hence we have proved that for any integers ${n\geq 18}$ and ${k \geq 2}$ such
  that\linebreak[4] ${\log_2 (2n-1)-2 \leq k \leq  n - 2}$, both conditions (2)
  and (3) are verified. Moreover, note that for any ${n\geq 18}$, there exists
  an integer $k \geq 2$ in the interval\linebreak[4]${\big[ \log_2 (2n-1)-2; n
  - 2 \big]}$. Indeed, ${\log_2 (2\cdot 18-1)-2 \approx 3.12 >2}$, the size
  of this interval increases with $n$, and it is larger than~$1$ for $n=18$.
  To conclude, remark that for such an integer $k$, condition (1) is easily
  verified from Theorem \ref{existdivnonspe} since $p^2\geq 4$ and ${g_k \geq
  g_2=3}$ according to Formula~(\ref{genregsr}).\\
  \qed
\end{Proof}

The similar result for the tower $T/\F_p$ is as follows:

\begin{lemm}\label{lemme_placedegngsr2}
  Let $p\geq5$ and $n \geq \frac{1}{2}\left(p+1+\epsilon(p)\right)$. There
  exists a step $H_k/\F_p$ of the tower $T/\F_p$ such that  the three following
  conditions are verified:
  \begin{enumerate}[(1)]
    \item there exists a non-special divisor of degree $g_k-1$ in $H_k/\F_p$,
    \item there exists a place of $H_k/\F_p$ of degree $n$,
    \item $N_1(H_k/\F_p) + 2N_1(H_k/\F_p) \geq 2n + 2g_k - 1$.
  \end{enumerate}
  Moreover, the first step for which all the three conditions are verified is
  the first step for which (3) is verified.
\end{lemm}

\begin{Proof}
  Note that ${n \geq \frac{1}{2}(5+1+\epsilon(5)) =5}$. We first prove that for
  all integers $k$ such that ${2 \leq k \leq n - 3}$, we have  ${2g_k+1 \leq
  p^\frac{n-1}{2}(\sqrt{p}-1)}$, so that condition (2) is verified according to
  Corollary 5.2.10 in \cite{stic2}. Indeed, for such an integer $k$, since
  ${p\geq5}$ and ${n\geq5}$ one has ${\log_2(p^\frac{n-1}{2}-1) \geq
  \log_2(5^\frac{n-1}{2}-1) \geq \log_2(2^{n-1}) = n-1}$.
  Thus\linebreak[4]${k+2\leq n-1\leq\log_2(p^\frac{n-1}{2}-1)}$ and it follows
  from Lemma \ref{lemme_genregsr}(ii) that \linebreak[4]${2g_k+1 \leq 2^{k+2}+1
  \leq p^\frac{n-1}{2} \leq p^\frac{n-1}{2}(\sqrt{p}-1)}$, which gives the
  result.

  The same reasoning as in the previous proof shows that condition (3) is also
  satisfied as soon as  ${k\geq \log_2 (2n-1)-2}$. Hence, we have proved that for
  any integers ${n\geq 5}$ and ${k \geq 2}$ such that ${\log_2 (2n-1)-2 \leq k
  \leq  n - 3}$, both conditions (2)\linebreak[4]and (3) are verified. Moreover,
  note that  the size of the interval\linebreak[4]${\big[ \log_2 (2n-1)-2; n - 3
  \big]}$ increases with $n$ and that for any ${n\geq 5}$, this interval contains
  at least one integer $k \geq 2$. To conclude, remark that for such an integer
  $k$, condition (1) is easily verified from Theorem \ref{existdivnonspe} since
  $p\geq 4$ and ${g_k \geq g_2=3}$ according to Formula (\ref{genregsr}).\\ \qed
\end{Proof}

Now we establish general bounds for the bilinear complexity of multiplication
by using derivative evaluations at places of degree one (respectively places of
degree one and two). The upcoming first theorem can be found in Arnaud's thesis
\cite{arna1}, but since the proof is rather short, we give it in order for this
article to be self-contained.
\begin{theo}{\label{thm_arnaud1}}
  Let $q$ be a prime power and $n>1$ be an integer. If there exists an
  algebraic function field $F/ \F_q$ of genus $g$ with $N$ places of degree 1
  and an integer $0 < a \leq N$ such that 
  \begin{enumerate}[(i)]
    \item there exists $\mathcal{R}$, a non-special divisor of degree $g-1$,
    \item there exists $Q$, a place of degree $n$,
    \item $N+a \geq 2n+2g-1$,
  \end{enumerate}
  then
  \begin{equation*}
    \mus_q(n) \leq 2n +g-1+a.
  \end{equation*}
\end{theo}

\begin{Proof}
  Let $\mathscr{P}:=\{P_1, \ldots, P_N\}$ be a set of $N$ places of degree 1
  and $\mathscr{P}'$ be a subset of $\mathscr{P}$ of cardinality $a$. According
  to Lemma 2.7 in \cite{bapi}, we can choose an effective divisor $\mathcal{D}$
  equivalent to $Q+\mathcal{R}$ such that ${\mathrm{supp}(\mathcal{D}) \cap
  \mathscr{P} = \varnothing}$. We define the maps $Ev_Q$ and $Ev_\mathscr{P}$
  as in Theorem \ref{theo_evalder} with $u_i=2$ if $P_i \in \mathscr{P}'$ and
  $u_i=1$ if $P_i \in \mathscr{P}\backslash\mathscr{P}'$. Then $Ev_Q$ is
  bijective, since $\ker Ev_Q = \mathcal{L}(\mathcal{D}-Q)$ with
  ${\dim(\mathcal{D}-Q) = \dim(\mathcal{R}) =0}$ and ${\dim (\mathrm{Im}\, Ev_Q
  )= \dim \mathcal{D} = \deg\mathcal{D} -g +1 + i(\mathcal{D}) \geq n}$
  according to the Riemann-Roch Theorem. Thus $\dim (\mathrm{Im}\, Ev_Q ) =n$.
  Moreover, $Ev_\mathscr{P}$ is injective. Indeed,\linebreak[4]${\ker
  Ev_\mathscr{P} = \mathcal{L}(2\mathcal{D}-\sum_{i=1}^N u_iP_i)}$ with $\deg
  (2\mathcal{D}-\sum_{i=1}^N u_iP_i) =2(n+g-1)-N-a <0$. Furthermore, one has
  $\mathrm{rank}\, Ev_\mathscr{P} = \dim(2\mathcal{D})=
  \deg(2\mathcal{D})-g+1+i(2\mathcal{D})$, and $i(2\mathcal{D})=0$ since
  $2\mathcal{D} \geq \mathcal{D} \geq \mathcal{R}$ with ${i(\mathcal{R})=0}$.
  So ${\mathrm{rank}\, Ev_\mathscr{P} = 2n+g-1}$, and we can extract a subset
  $\mathscr{P}_1$ of $\mathscr{P}$ and a subset $\mathscr{P}_1'$ of
  $\mathscr{P}'$ with cardinality $N_1\leq N$ and $a_1\leq a$, such that:
  \begin{itemize}
    \item $N_1+a_1 = 2n+g-1$,
    \item the map $Ev_{\mathscr{P}_1}$ defined as $Ev_\mathscr{P}$ with $u_i=2$
      if ${P_i \in \mathscr{P}_1'}$ and $u_i=1$ if ${P_i \in
        \mathscr{P}_1\backslash\mathscr{P}_1'}$, is injective.
  \end{itemize}
  According to Theorem \ref{theo_evalder}, this leads to $\mu_q(n) \leq
  N_1+2a_1 \leq N_1+a_1+a$, which gives the result.
\qed\\
\end{Proof}

This second theorem is a refinement of \cite[Theorem 3.8]{arna1}, that will
allow us to improve Arnaud's bound for $\mus_q(n)$ and $\mus_p(n)$ in the next
paragraph.

\begin{theo}{\label{thm_arnaud2}}
  Let $q>2$ be a prime power and $n>1$ be an integer. If there exists an
  algebraic function field $F/ \F_q$ of genus $g$ with $N_1$ places of degree
  1, $N_2$ places of degree 2, and two integers  $0 < a_1 \leq N_1$, $0 < a_2
  \leq N_2$ such that 
  \begin{enumerate}[(i)]
    \item there exists  $\mathcal{R}$, a non-special divisor of degree $g-1$,
    \item there exists $Q$, a place of degree $n$,
    \item $N_1+a_1 +2(N_2+a_2) \geq 2n+2g-1$,
  \end{enumerate}
  then
  \begin{equation*}
    \mus_q(n) \leq 2n + g +N_2 + a_1 + 4a_2
  \end{equation*}
  and
  \begin{equation*}
    \mus_q(n) \leq 3n+2g+\frac{a_1}{2}+3a_2. 
  \end{equation*}
\end{theo}

\Remark Under the same hypotheses, the bounds obtained in
\cite[Theorem~3.8]{arna1} are
$
{\mus_q(n) \leq 2n + 2g +N_2 + a_1 + 4a_2}
$
and
$
{\mus_q(n) \leq 3n+3g+\frac{a_1}{2}+3a_2.} 
$

\begin{Proof}
  We use the same notation as in Corollary \ref{theo_deg12evalder}: ${\mathscr
  P := \{P_1,\ldots,P_{N_1}\}}$ is a set of $N_1$ places of degree one and
  ${\mathscr{P}' := \{R_{1},\ldots,R_{N_2}\}}$ is a set of $N_2$ places of
  degree two. According to hypothesis (iii), one can always reduce to the
  case where
  \begin{equation}\label{2n2g}
    2n+2g-1\leq N_1+a_1 +2(N_2+a_2) \leq 2n+2g.
  \end{equation}
  According to Lemma 2.7 in \cite{bapi}, we can choose an effective divisor
  $\mathcal{D}$ equivalent to $Q+\mathcal{R}$ such that
  ${\mathrm{supp}(\mathcal{D}) \cap (\mathscr{P}\cup \mathscr{P}') =
  \varnothing}$. We then define the maps $Ev_Q$ and $Ev_{\mathscr P, \mathscr
  P'}$  as in Corollary \ref{theo_deg12evalder} but for the second one, we fix
  $\F_q^{N_1+a_1+2(N_2+a_2)}$  as codomain instead of ${\F_{q}^{N_1} \times
  \F_{q}^{a_1}\times \F_{q^2}^{N_2} \times \F_{q^2}^{a_2}}$ (this means that we
  choose a basis of $\F_{q^2}$ over $\F_q$ and take the components of each
  element of $\F_{q^2}$ with respect to this basis).
  
  The same reasoning as in the previous proof shows that $Ev_Q$ is bijective.
  Moreover, the map $Ev_{\mathscr P, \mathscr P'}$  is injective since
  \begin{equation*}
    {\ker Ev_\mathscr{P} = \mathcal{L}\left(2\mathcal{D}-\left(\sum_{i=1}^{N_1}
    P_i + \sum_{i=1}^{a_1} P_i+\sum_{i=1}^{N_2} R_i + \sum_{i=1}^{a_2}
    R_i\right)\right)}
  \end{equation*}
  with $\deg \left(2\mathcal{D}-\left(\sum_{i=1}^{N_1} P_i + \sum_{i=1}^{a_1}
  P_i+\sum_{i=1}^{N_2} R_i + \sum_{i=1}^{a_2} R_i\right)\right)<0$ from
  hypothesis (iii).  Furthermore, one has $\mathrm{rank}\,
  Ev_{\mathscr{P},\mathscr{P}'} = \dim(2\mathcal{D})=
  \deg(2\mathcal{D})-g+1+i(2\mathcal{D})$, and $i(2\mathcal{D})=0$ since
  $2\mathcal{D} \geq \mathcal{D} \geq \mathcal{R}$ with ${i(\mathcal{R})=0}$.
  So ${\mathrm{rank}\, Ev_{\mathscr{P, P'}} = 2n+g-1}$. Thus, $Ev_{\mathscr{P,
  P'}}$ being injective with rank $2n+g-1$, it follows that one can choose a
  suitable subset of coordinates of size ${2n+g-1}$ (among the
  ${N_1+a_1+2(N_2+a_2)}$ ones in $\F_q^{N_1+a_1+2(N_2+a_2)}$) of any element in
  the image to define its preimage.

  Now we will focus on the number of multiplications in $\F_q$ needed to define
  the ${2n+g-1}$ coordinates of the image of a product $fg$ for $f,g\in
  \Ld{2}$, from the coordinates of the images of $f$ and $g$. Note that we will
  need more than the two subsets of $2n+g-1$ coordinates from $Ev_{\mathscr{P,
  P'}}(f)$ and $Ev_{\mathscr{P, P'}}(g)$ to compute the coordinates of the
  image $fg$. But in the end, we need only ${2n+g-1}$ of these coordinates to
  define the preimage of $fg$ in $\Ld{2}$. There are 4 types of such ``useful''
  coordinates:
  \begin{enumerate}[(a)]
    \item those which come from a classical evalution over a place of degree 1,
      such as $f(P_1)$; we denote the number of such coordinates by $L_1$.
    \item those which come from a derivated evalution over a place of degree 1,
      such as $f'(P_1)$; we denote the number of such coordinates by $\ell_1$.
    \item those which come from a classical evalution over a place of degree 2,
      such as both coordinates in $\F_q$ of $f(R_1)$; we denote the number of
      such coordinates by $L_2$.
    \item those which come from a derivated evalution over a place of degree 2,
      such as both coordinates in $\F_q$ of $f'(R_1)$; we denote the number of
      such coordinates by $\ell_2$.
  \end{enumerate}
  With these notations, we have that:
  \begin{equation}\label{2ng}
    L_1+\ell_1+L_2+\ell_2 = 2n + g -1
  \end{equation}
  with 
  \begin{equation}\label{Lili}
    \mbox{$L_1 \leq N_1$, \quad $\ell_1 \leq a_1$, \quad $L_2 \leq 2N_2$ \quad
    and \quad $\ell_2 \leq 2a_2$.}\\
  \end{equation}

  Now we will estimate how many multiplications in $\F_q$ are needed to compute
  each type of coordinate for the image of the product $fg$.
  \begin{itemize}
    \item to obtain a type (a) coordinate, we need 1 multiplication in $\F_q$
      since
      \begin{equation*}
        (fg)(P_i) = f(P_i)\cdot g(P_i) \quad \mbox{with } f(P_i),g(P_i) \in
        \Fq.
      \end{equation*}
    \item to obtain a type (b) coordinate, we need 2 multiplications in $\F_q$
      since
      \begin{equation*}
        (fg)'(P_i) = f'(P_i) \cdot g(P_i) +f(P_i) \cdot g'(P_i) \quad
        \mbox{with } f(P_i),g(P_i), f'(P_i),g'(P_i) \in \Fq.
    \end{equation*}
    \item to obtain a type (c) coordinate, we need 2 multiplications in $\F_q$.
      Indeed, this type of coordinate is only one of the two coordinates in
      $\F_q$ of an element in $\F_{q^2}$. For example, if we denote by
      $\big(f(R_i)_1,f(R_i)_2\big)$ the two coordinates in $\F_q$ of the vector
      which represents ${f(R_i)\in \F_{q^2}}$, in a $\F_q$-basis ${(1,
      \alpha)}$ of $\F_{q^2}$ where ${\alpha^2 =-1}$, we get
      \begin{equation*}
        (fg)(R_i)_1 = f(R_i)_1 \cdot g(R_i)_1 - f(R_i)_2 \cdot g(R_i)_2
      \end{equation*}
      for the first coordinate, or
      \begin{equation*}
        (fg)(R_i)_2 = f(R_i)_1 \cdot g(R_i)_2 + f(R_i)_2 \cdot g(R_i)_1
      \end{equation*}
      for the second one.
    \item to obtain a type (d) coordinate, we need 4 multiplications in $\F_q$
      since we have to determine either $U \in \F_q$ or $V \in \F_{q}$ such
      that: 
      \begin{equation*}
        (fg)'(R_i) =  f'(R_i)\cdot g(R_i) + f(R_i) \cdot g'(R_i) = U+ \alpha V
      \end{equation*}
      so we need to compute
      \begin{equation*}
        U = f(R_i)_1 \cdot g'(R_i)_1 - f(R_i)_2 \cdot g'(R_i)_2 + g(R_i)_1
        \cdot f'(R_i)_1 - g(R_i)_2 \cdot f'(R_i)_2
      \end{equation*}
      or
      \begin{equation*}
        V = f(R_i)_2 \cdot g'(R_i)_1 + f(R_i)_1 \cdot g'(R_i)_2 + g(R_i)_2
        \cdot f'(R_i)_1 + g(R_i)_1 \cdot f'(R_i)_2.
      \end{equation*}
  \end{itemize}

  So far, it seems that we need ${L_1+2\ell_1+2L_2+4\ell_2}$ multiplications in
  $\F_q$ to obtain the  ${L_1+\ell_1+L_2+\ell_2 = 2n+g-1}$ coordinates of a
  product, which would be bounded by ${N_1+2a_1+4N_2+8a_2}$ according to
  (\ref{Lili}). We have to be a bit more precise to obtain a better bound.
  Indeed, when we use more than half the coordinates in $\F_q$ coming from
  places of degree 2, we know that we can be more efficient since we will have
  to compute some coordinates which come from the same evaluation. Namely, if
  we know that we will have to compute both $(fg)(R_i)_1$ and $(fg)(R_i)_2$ for
  some $i$, then we would not need $2 \cdot 2 = 4$ multiplications in $\F_q$,
  but only 3, thanks to Karatsuba algorithm. The same reasonning holds for
  derivated evalutations at places of degree 2: if we need to compute both
  $(fg)'(R_i)_1$ and $(fg)'(R_i)_2$, then we would not need $2 \cdot 4 = 8$
  multiplications in $\F_q$ but only 6.

  We therefore have to distinguish cases were we know how many ``paired''
  coordinates we have. Here is how we proceed:
  \begin{center}
    \begin{tabular}{ | c |c | c |}
      \hline
      & $L_2 \leq N_2$ & $N_2 < L_2 \leq 2N_2$ \\
      \hline
      $\ell_2 \leq a_2$ & Case 1 & Case 2 \\
      \hline
      $a_2 < \ell_2 \leq 2a_2$ & Case 3 & Case 4 \\
      \hline
    \end{tabular}
  \end{center}
  Thus, for the $L_2$ type (c) coordinates, we know that in cases 1 and 3,
  there are at least ${2(L_2 - N_2)}$ ``paired'' coordinates (since ${L_2 \leq
  N_2}$), and that each couple requires 3 multiplications in $\F_q$, so we
  perform ${3(L_2 - N_2)}$ such multiplications. The remaining ${2N_2-L_2}$
  coordinates have to be computed independently: it costs 2 multiplications in
  $\F_q$ for each.

  The same reasoning applies to the type (d) coordinates in cases 2 and 4:
  since ${N_2 < L_2 \leq 2N_2}$, there are ${2(L_2 - N_2)}$ coordinates which
  can be computed ``pairwise'', each couple needing 3 multiplications in
  $\F_q$, so we perform ${3(L_2 - N_2)}$ multiplications in $\F_q$. The
  remaining ${2N_2-L_2}$ coordinates have to be computed independently; it
  costs 4 multiplications in $\F_q$ for each.

  From this reasoning and the inequalities  (\ref{2ng}) and  (\ref{Lili}), we
  get the following bounds for the obtention of the ${2n+g-1}$ coordinates of a
  product:

  \noindent \textbf{Case 1:}
  \begin{eqnarray*}
    L_1+2\ell_1+2L_2+4\ell_2 & = & (L_1+\ell_1+L_2+\ell_2) +\ell_1+L_2+3\ell_2
      \\
    & \leq & 2n+g-1+a_1+N_2+3a_2
  \end{eqnarray*}
  \noindent \textbf{Case 2:} 
  \begin{eqnarray*}
    L_1+2\ell_1+3(L_2-N_2)+2(2N_2-L_2)+4\ell_2 & = & L_1+2\ell_1+N_2+
      L_2+4\ell_2\\
    & = & (L_1+\ell_1+L_2+\ell_2) +\ell_1+N_2+3\ell_2 \\
    & \leq & 2n+g-1+a_1+N_2+3a_2
  \end{eqnarray*}
  \noindent \textbf{Case 3:} 
  \begin{eqnarray*}
    L_1+2\ell_1+2L_2+6(\ell_2-a_2) + 4(2a_2- \ell_2) & = &  L_1+2\ell_1 +2L_2
      +2(a_2+ \ell_2) \\
    & = & (L_1+\ell_1+L_2+\ell_2) +\ell_1+L_2+\ell_2+2a_2 \\
    & \leq & 2n+g-1+a_1+N_2+4a_2
  \end{eqnarray*}
  \noindent \textbf{Case 4:} 
  \begin{eqnarray*}
    & & L_1+2\ell_1+3(L_2-N_2)+2(2N_2-L_2)+6(\ell_2-a_2) + 4(2a_2 -\ell_2)  \\
    & = & L_1+2\ell_1+L_2+N_2+2(a_2+\ell_2) \\
    & = & (L_1+\ell_1+L_2+\ell_2) + \ell_1 +N_2+\ell_2+2a_2\\
    & \leq & 2n+g-1+a_1+N_2+4a_2\\
  \end{eqnarray*}
  Thus $2n+g-1+a_1+N_2+4a_2$ is a bound which holds in all the four cases, so
  it gives an upper bound for the minimal number of multiplications in $\F_q$
  needed to obtain the ${2n+g-1}$ coordinates in $\F_q$ necessary to define a
  preimage by $Ev_{\mathscr{P}, \mathscr{P}'}$ of an element ${Ev_{\mathscr{P},
  \mathscr{P}'}(fg) \in \F_q^{N_1+a_1+2(N_2+a_2)}}$. Thus we have that
  \begin{equation*}
    \mus_q(n) \leq 2n+g-1+a_1+N_2+4a_2. 
  \end{equation*}
  The second bound of the theorem comes from (\ref{2n2g}), which implies
  that\linebreak[4]${{a_1 \over 2} + N_2+a_2 \leq n+g}$, and therefore
  \begin{equation*}
    2n+g-1+a_1+N_2+4a_2 \leq 3n +2g + {a_1 \over 2} + 3 a_2.
  \end{equation*}
  \qed
\end{Proof}

\subsection{Proof of the upper bounds stated in the
introduction}\label{sectbornesarnaud}

Here we give the detailed proof of Theorems \ref{theo_arnaudupdate} and
\ref{theo_arnaud1} by combining the results of the previous section. We use the
same notations concerning the number of places and the genera of curves in the
towers. Recall that depending on the tower under consideration the following
holds:

\begin{itemize}
  \item ${N_{k,s} := N_1(F_{k,s}/\F_{q^2}) = N_{k,s} =
    N_1(G_{k,s}/\F_q)+2N_2(G_{k,s}/\F_q)}$
  \item ${N_k:= N_1(H_k/\F_{p^2})=N_1(H_k/\F_p)+2N_2(H_k/\F_p)}$
  \item ${\Delta g_{k,s} := g_{k,s+1} - g_{k,s}}$ and ${\Delta g_{k} := g_{k+1}
    - g_{k}}$
  \item ${D_{k,s}:=(p-1)p^sq^k}$. 
\end{itemize}

\begin{Proofof}[Theorem \ref{theo_arnaudupdate}]
  \begin{enumerate}
    \item[\textbf{(i)}] Let $n\geq \frac{1}{2}(q+1+{\epsilon (q) })$; in the
      complementary case, we already know from Section \ref{sect_known} that
      $\mus_{q}(n) \leq 2n$. According to Lemma \ref{lemme_placedegn2}, there
      exists a step of the tower $T_3/\F_{q}$ to which we can apply Theorem
      \ref{thm_arnaud2} with $a_1=a_2=0$. We denote by $G_{k,s+1}/\F_{q}$ the
      first step of the tower that satisfies the hypotheses of Theorem
      \ref{thm_arnaud2} with $a_1=a_2=0$, i.e. $k$ and $s$ are
      integers\linebreak[4] such that ${N_{k,s+1} \geq 2n+2g_{k,s+1}-1}$ and
      ${N_{k,s}< 2n+2g_{k,s}-1}$, where \linebreak[4]${N_{k,s} :=
      N_1(G_{k,s}/\F_q)+2N_2(G_{k,s}/\F_q)}$ and ${g_{k,s} := g(G_{k,s})}$. We
      denote by $n_0^{k,s}$ the biggest integer such that ${N_{k,s}\geq
      2n_0^{k,s} + 2g_{k,s}-1}$, so that we have the equality ${n_0^{k,s} =
      \sup \big\{n \in \N \, \vert \, 2n \leq N_{k,s} - 2g_{k,s} + 1\big\}}$.
      To perform multiplication in $\F_{q^n}$, we have the following
      alternative approaches:
      \begin{enumerate}[(a)]
      \item use the algorithm at step $G_{k,s+1}$. In this case, a bound for
        the bilinear complexity is given by Theorem \ref{thm_arnaud2} applied
        with $a_1 = a_2 = 0$:
        \begin{equation*}
          \mus_q(n) \leq 3n+2g_{k,s+1} = 3n_0^{k,s}+2g_{k,s}+3(n-n_0^{k,s}) +
          2\Delta g_{k,s}.
        \end{equation*}
      \item use the algorithm on the step $G_{k,s}$ with an appropriate number
        of derivative evaluations. Let ${a_1+2a_2:= 2(n-n_0^{k,s})}$ then
        ${N_{k,s}\geq 2n_0^{k,s}+2g_{k,s}-1}$, implies that ${N_{k,s} +a_1+2a_2
        \geq 2n+2g_{k,s}-1}$. Thus , if ${a_1+2a_2 \leq N_{k,s}}$, we can
        perform $a_1+a_2$ derivative evaluations in the algorithm using the
        step $G_{k,s}$ and we have:
        \begin{equation*}
          \mus_q(n) \leq 3n+2g_{k,s} + \frac{3}{2}(a_1+2a_2)
          = 3n_0^{k,s}+2g_{k,s}+6(n-n_0^{k,s}).
        \end{equation*}
      \end{enumerate}
      Thus if $a_1+2a_2 \leq N_{k,s}$, then case (b) gives a better bound as
      soon as\linebreak[4]${n-n_0^{k,s}<\frac{2}{3}\Delta g_{k,s}}$. So we have
      from Lemma \ref{lemme_delta}, with ${\Tilde{D}_{k,s} :=
      \frac{3}{4}D_{k,s}}$:\linebreak[4] ${N_{k,s} \geq
      \frac{4}{3}\Tilde{D}_{k,s}}$ and ${\Delta g_{k,s} \geq \Tilde{D}_{k,s}}$.
      Hence if $a_1+2a_2< \frac{4}{3} \Tilde{D}_{k,s}$ (i.e.
      \linebreak[5]${2(n-n_0^{k,s})< \frac{4}{3}\Tilde{D}_{k,s}}$), then we
      both have that ${2(n-n_0^{k,s}) < \frac{4}{3}\Delta g_{k,s}}$
      and\linebreak[4] ${a_1+2a_2 \leq N_{k,s}}$. We can therefore perform
      $a_1$ derivative evaluations at places of degree 1 and $a_2$ derivative
      evaluations at places of degree 2 in the step $G_{k,s}$ and case (b)
      gives a better bound than case (a). Moreover, $a_1+2a_2 < \frac{4}{3}
      \Tilde{D}_{k,s}$ is equivalent to ${n-n_0^{k,s} < D_{k,s}}$.

      For $x \in \mathbb{R}^{+}$ such that ${N_{k,s+1} \geq 2[x]+2g_{k,s+1}-1}$
      and \linebreak[4]${N_{k,s} < 2[x]+2g_{k,s}-1}$, we define the function
      $\Phi_{k,s}(x)$ as follows:
      \begin{equation*}
        \Phi_{k,s}(x) = \left\{
          \begin{array}{ll}
            3x+2g_{k,s}+3(x-n_0^{k,s}) & \mbox{if } x- n_0^{k,s} < D_{k,s}\\
            & \\
            3x+2g_{k,s+1}& \mbox{otherwise}.
          \end{array}
        \right.
      \end{equation*}
      We define the function $\Phi$ for ${x\geq0}$ to be the minimum of the
      functions $\Phi_{k,s}$ for which $x$ is in the domain of $\Phi_{k,s}$.
      This function is piecewise linear, with two kinds of pieces: those which
      have slope $3$ and those which have slope~$6$. Moreover, since the
      $y$-intercept of each piece grows with $k$ and $s$, the graph of the
      function $\Phi$ lies below any straight line that lies above all the
      points ${\big(n_0^{k,s} + D_{k,s}, \Phi(n_0^{k,s} + D_{k,s})\big)}$,
      since these are the vertices of the graph. If we let ${X:=n_0^{k,s} +
      D_{k,s}}$, then
      \begin{eqnarray*}
        \Phi(X) & \leq & 3X + 2g_{k,s+1} \\
        & = & 3\left(1 + \frac{2g_{k,s+1}}{3X}\right)X.
      \end{eqnarray*}
      We want to give a bound for $\Phi(X)$ that is independent of $k$ and $s$.
      Recall that $D_{k,s} :=(p-1)p^sq^k$, and
      \begin{equation*}
        n_0^{k,s}  \geq \frac{1}{2}q^{k-1}p^s(q+1)(q-3) \  \  \  \mbox{by Lemma
          \ref{lemme_bornesup}}
      \end{equation*}
      and
      \begin{equation*}
        g_{k,s+1} \leq q^{k-1}(q+1)p^{s+1}  \  \  \  \mbox{by Lemma
          \ref{lemme_genre} (iii).}
      \end{equation*}
      So we have
      \begin{eqnarray*}
        \frac{2g_{k,s+1}}{3X} & = & \frac{2g_{k,s+1}}{3(n_0^{k,s}+D_{k,s})} \\
        & \leq & \frac{2q^{k-1}(q+1)p^{s+1}}{3(\frac{1}{2}q^{k-1}p^s(q+1)(q-3)
          + (p-1)p^sq^k)} \\
        & = & \frac{2q^{k-1}(q+1)p^sp}{q^{k-1}(q+1)p^s\left(\frac{3}{2}(q-3) +
          3(p-1)\frac{q}{q+1}\right) }\\
        & = & \frac{\frac{4}{3}p}{(q-3)+2(p-1)\frac{q}{q+1}}.
      \end{eqnarray*}
      Thus the graph of the function $\Phi$ lies below the line ${y=3\left(1 +
      \frac{\frac{4}{3}p}{(q-3)+2(p-1)\frac{q}{q+1}}\right)x}$. In particular,
      we obtain
      \begin{equation*}
        \Phi(n) \leq 3\left(1 + \frac{\frac{4}{3}p}{(q-3) +
          2(p-1)\frac{q}{q+1}}\right)n.
      \end{equation*}

    \item[\textbf{(ii)}] Let $n\geq \frac{1}{2}(p+1+{\epsilon (p) })$; in the
      complementary case, we already know from Section \ref{sect_known} that
      $\mus_{p}(n) \leq 2n$. According to Lemma \ref{lemme_placedegngsr2},
      there exists a step of the tower $T/\F_p$ on which we can apply Theorem
      \ref{thm_arnaud2} with ${a_1=a_2=0}$. We denote by $H_{k+1}/\F_p$ the first
      step of the tower that satisfies the hypotheses of Theorem
      \ref{thm_arnaud2} with $a_1=a_2=0$, i.e. $k$ is an integer such that\linebreak[4]
      ${N_{k+1} \geq 2n+2g_{k+1}-1}$ and ${N_k< 2n+2g_k-1}$, where\linebreak[4]
      ${N_k:=N_1(H_k/\F_p)+2N_2(H_k/\F_p)}$ and ${g_k:=g(H_k)}$. We denote by
      $n_0^k$ the biggest integer such that we have ${N_k\geq 2n_0^k+2g_k-1}$,
      i.e. \linebreak[4]${n_0^k = \sup \big\{n \in \N \, \vert \, 2n \leq
      N_k-2g_k+1\big\}}$. To perform multiplication in $\F_{p^n}$, we have the
      following alternative approaches:
      \begin{enumerate}[(a)]
        \item use the algorithm at the step $H_{k+1}$. In this case, a bound for
          the bilinear complexity is given by Theorem \ref{thm_arnaud2} applied
          with $a_1=a_2=0$:
          \begin{equation*}
            \mus_q(n) \leq 3n+2g_{k+1}= 3n_0^k+2g_k+3(n-n_0^k) + 2\Delta g_k.
          \end{equation*}
        \item use the algorithm at the step $H_k$ with an appropriate number of
          derivative evaluations. If we let ${a_1+2a_2:= 2(n-n_0^k)}$, then
          ${N_k\geq 2n_0^k+2g_k-1}$ implies that ${N_k +a_1+2a_2 \geq
          2n+2g_k-1}$. Thus if ${a_1+2a_2 \leq N_k}$, we can perform $a_1+a_2$
          derivative evaluations in the algorithm using the step $H_k$, and we
          have:
          \begin{equation*}
            \mus_p(n) \leq 3n+2g_k+\frac{3}{2}(a_1+2a_2)=3n_0^k+2g_k+6(n-n_0^k).
          \end{equation*}
      \end{enumerate}
      Thus, if $a_1+2a_2 \leq N_{k,s}$, then case (b) gives a better bound as
      soon\linebreak[4]as ${n-n_0^{k,s}<\frac{2}{3}\Delta g_{k,s}}$.
    
      For $x \in \mathbb{R}^{+}$ such that ${N_{k+1} \geq 2[x]+2g_{k+1}-1}$ and
      ${N_k < 2[x]+2g_k-1}$, we define the function $\Phi_k(x)$ as follows:
      \begin{equation*}
        \Phi_k(x) = \left\{
          \begin{array}{ll}
            3x+2g_k+3(x-n_0^k) & \mbox{if } x- n_0^k < \frac{3}{2}\Delta g_k\\
            & \\
            3x+2g_{k+1}& \mbox{otherwise}.
          \end{array}
        \right.
      \end{equation*}
      Note that when case (b) gives a better bound, that is to say
      when\linebreak[4]${\frac{3}{2}(x-n_0^k) < \Delta g_k}$, then according to
      Lemma \ref{lemme_deltagsr} we also have that
      \begin{equation*}
        2(x-n_0^k)< N_k
      \end{equation*}
      since ${\frac{4}{3}\Delta g_k \leq N_k}$. We can therefore proceed as in
      case (b), since there are enough places of degree 1 and 2 at which we can
      perform $a_1+a_2=2(x-n_0^k)$ derivative evaluations on.

      We define the function $\Phi$ for ${x\geq0}$ to be the minimum of the
      functions $\Phi_k$ for which $x$ is in the domain of $\Phi_k$. This
      function is piecewise linear with two kinds of pieces: those which have
      slope $3$ and those which have slope~$6$. Moreover, since the
      $y$-intercept of each piece grows with $k$, the graph of the function
      $\Phi$ lies below any straight line that lies above all the
      points\linebreak[4]${\big(n_0^k+\frac{2}{3}\Delta g_k,
      \Phi(n_0^k+\frac{2}{3}\Delta g_k)\big)}$, since these are the vertices of
      the graph. If we let ${X:=n_0^k+\frac{2}{3}\Delta g_k}$, then
      \begin{eqnarray*}
        \Phi(X) & \leq & 3X + 2g_{k+1} = 3\left(1 +
        \frac{2g_{k+1}}{3X}\right)X.
      \end{eqnarray*}
      We want to give a bound for $\Phi(X)$ that is independent of $k$. Lemmas
      \ref{lemme_genregsr}(ii), \ref{lemme_deltagsr} and
      \ref{lemme_bornesupgsr} give:
      
      \begin{eqnarray*}
        \frac{2g_{k+1}}{3X}  & \leq & \frac{2^{k+3}}{3\left(2^{k}(p-3) + 2 +
          \frac{2}{3}(2^{k+1} - 2^{\frac{k+1}{2}})\right)} \\
        & = & \frac{8\cdot 2^{k}}{2^k\left( 3(p-3) + 3\cdot 2^{-k+1} + 4 (1 -
          2^{-\frac{k+1}{2}})\right)}\\
        & = & \frac{8/3}{p-3+\frac{4}{3} + 2^{-k+1} - \frac{1}{3}
          2^{-\frac{k-3}{2}}}\\
        & \leq & \frac{8/3}{p-\frac{5}{3}}
      \end{eqnarray*}
      since ${2^{-k+1}-\frac{1}{3}2^{-\frac{k-3}{2}}\geq 0}$. Thus the graph
      of the function $\Phi$ lies below the line ${y=3\left(1 + \frac{8}{3p-5}
      \right)x}$. In particular, we obtain
      \begin{equation*}
        \Phi(n) \leq 3\left(1+ \frac{8}{3p-5}\right)n.
      \end{equation*}
    \end{enumerate}
  \qed
\end{Proofof}

\begin{Proofof}[Theorem  \ref{theo_arnaud1}]
  \begin{enumerate}[(i)]
    \item[\textbf{(i)}] Let $n\geq \frac{1}{2}(q^2+1+{\epsilon (q^2) })$; in
      the complementary case, we already know from the pioneering works
      recalled in Section \ref{sect_known} that $\mus_{q^2}(n) \leq 2n$.
      According to Lemma \ref{lemme_placedegn}, there exists a step of the
      tower $T_2/\F_{q^2}$ at which we can apply Theorem \ref{thm_arnaud1} with
      $a=0$. We denote by $F_{k,s+1}/\F_{q^2}$ the first step of the tower that
      satisfies the hypotheses of Theorem \ref{thm_arnaud1} with $a=0$, i.e. $k$
      and $s$ are integers such that ${N_{k,s+1} \geq 2n+2g_{k,s+1}-1}$ and
      ${N_{k,s}< 2n+2g_{k,s}-1}$, where ${N_{k,s}:=N_1(F_{k,s}/\F_{q^2})}$ and
      ${g_k:=g(F_{k,s})}$. We denote by $n_0^{k,s}$ the biggest integer such
      that ${N_{k,s}\geq 2n_0^{k,s}+2g_{k,s}-1}$, i.e. ${n_0^{k,s}
      = \sup \big\{n \in \N \, \vert \, 2n \leq N_{k,s}-2g_{k,s}+1\big\}}$. To
      perform multiplication in $\F_{q^{2n}}$, we have the following
      alternative approaches:
      \begin{enumerate}[(a)]
        \item use the algorithm at the step $F_{k,s+1}$. In this case, a bound
          for the bilinear complexity is given by Theorem \ref{thm_arnaud1}
          applied with $a=0$:
          \begin{equation*}
            \mus_{q^2}(n) \leq 2n+g_{k,s+1}-1 = 2n+g_{k,s}-1 +\Delta g_{k,s}.
          \end{equation*}
          (Recall that $\Delta g_{k,s} := g_{k,s+1} - g_{k,s}$.)
        \item use the algorithm at the step $F_{k,s}$ with an appropriate
          number of derivative evaluations. Let $a:= 2(n-n_0^{k,s})$ and
          suppose that $a \leq N_{k,s}$. Then ${N_{k,s} \geq
          2n_0^{k,s}+2g_{k,s}-1}$ implies that ${N_{k,s} +a \geq
          2n+2g_{k,s}-1}$, so condition (iii) of Theorem \ref{thm_arnaud1} is
          satisfied. Thus, we can perform $a$ derivative evaluations in the
          algorithm using the step $F_{k,s}$ and we have:
          \begin{equation*}
            \mus_{q^2}(n) \leq 2n+g_{k,s}-1+a.
          \end{equation*}
      \end{enumerate}
      Thus, if $a \leq N_{k,s}$, then case (b) gives a better bound as soon as
      ${a<\Delta g_{k,s}}$. Since Lemma \ref{lemme_delta} gives the
      inequalities ${N_{k,s} \geq D_{k,s}}$ and ${\Delta g_{k,s} \geq
      D_{k,s}}$, we know that if ${a\leq D_{k,s}}$, then we can perform $a$
      derivative evaluations on places of degree 1 in the step $F_{k,s}$. This
      implies that case (b) gives a better bound than case (a).

      For $x \in \mathbb{R}^{+}$ such that ${N_{k,s+1} \geq 2[x]+2g_{k,s+1}-1}$
      and ${N_{k,s} < 2[x]+2g_{k,s}-1}$, we define the function $\Phi_{k,s}(x)$
      as follows:
      \begin{equation*}
        \Phi_{k,s}(x) = \left\{
          \begin{array}{ll}
            2x+g_{k,s}-1+2(x-n_0^{k,s}) & \mbox{if } 2(x- n_0^{k,s}) <
              D_{k,s}\\
            2x+g_{k,s+1}-1& \mbox{else}.
          \end{array}
        \right.
      \end{equation*}
      We define the function $\Phi$ for ${x\geq0}$ to be the minimum of the
      functions $\Phi_{k,s}$ for which $x$ is in the domain of $\Phi_{k,s}$.
      This function is piecewise linear with two kinds of pieces: those which
      have slope $2$ and those which have slope~$4$. Moreover, since the
      $y$-intercept of each piece grows with $k$ and $s$, the graph of the
      function $\Phi$ lies below any straight line that lies above all the
      points ${\big(n_0^{k,s}+\frac{D_{k,s}}{2}, \Phi(n_0^{k,s} +
      \frac{D_{k,s}}{2})\big)}$, since these are the vertices of the
      graph. If we let ${X:=n_0^{k,s} + \frac{D_{k,s}}{2}}$, then we have
      \begin{eqnarray*}
        \Phi(X) & \leq & 2X + g_{k,s+1} -1\\
        & \leq & 2X+ g_{k,s+1}\\
        & = & 2\left(1 + \frac{g_{k,s+1}}{2X}\right)X.
      \end{eqnarray*}
      We want to give a bound for $\Phi(X)$ which is independent of $k$ and
      $s$.

      Recall that $D_{k,s} := (p-1) p^s q^k$, and
      \begin{equation*}
        2n_0^{k,s}  \geq q^{k-1}p^s(q+1)(q-3) \  \  \  \mbox{by Lemma
          \ref{lemme_bornesup}}
      \end{equation*}
      and
      \begin{equation*}
        g_{k,s+1} \leq q^{k-1}(q+1)p^{s+1}  \  \  \  \mbox{by Lemma
          \ref{lemme_genre} (iii).}
      \end{equation*}
      So we have
      \begin{eqnarray*}
        \frac{g_{k,s+1}}{2X} & = & \frac{g_{k,s+1}}{2n_0^{k,s}+D_{k,s}} \\
        & \leq & \frac{q^{k-1}(q+1)p^{s+1}}{q^{k-1}p^s(q+1)(q-3) + (p-1)p^sq^k}
      \end{eqnarray*}
      \begin{eqnarray*}
        \frac{g_{k,s+1}}{2X} & \leq & \frac{q^{k-1}(q+1) p^s p}{q^{k-1} (q+1)
          p^s \left(q-3 + (p-1)\frac{q}{q+1} \right) }\\
        & = & \frac{p}{(q-3)+(p-1)\frac{q}{q+1}}.
      \end{eqnarray*}
      Thus, the graph of the function $\Phi$ lies below the line ${y=2\left(1 +
      \frac{p}{(q-3)+(p-1)\frac{q}{q+1}}\right)x}$. In particular, we obtain
      \begin{equation*}
        \Phi(n) \leq 2\left(1 + \frac{p}{(q-3)+(p-1)\frac{q}{q+1}}\right)n.
      \end{equation*}

    \item[\textbf{(ii)}] Let $n\geq \frac{1}{2}(p^2+1+{\epsilon (p^2) })$;
      in the complementary case, we already know from Section \ref{sect_known}
      that $\mus_{p^2}(n) \leq 2n$. According to Lemma
      \ref{lemme_placedegngsr}, there exists a step of the tower $T/\F_{p^2}$
      at which we can apply Theorem \ref{thm_arnaud1} with $a=0$. We denote by
      $H_{k+1}/\F_{p^2}$ the first step of the tower that satisfies the
      hypotheses of Theorem~\ref{thm_arnaud1} with $a=0$, i.e. $k$ is an
      integer such that ${N_{k+1} \geq 2n+2g_{k+1}-1}$ and ${N_k< 2n+2g_k-1}$,
      where ${N_k:=N_1(H_k/\F_{p^2})}$ and ${g_k:=g(H_k)}$. We denote by
      $n_0^k$ the biggest integer such that ${N_k\geq 2n_0^k+2g_k-1}$, so that
      we have the equality ${n_0^k = \sup \big\{n \in \N \, \vert \, 2n \leq
      N_k-2g_k+1\big\}}$. To perform multiplication in $\F_{p^{2n}}$, we have
      the following alternative approaches:
      \begin{enumerate}[(a)]
        \item use the algorithm at the step $H_{k+1}$. In this case, a bound
          for the bilinear complexity is given by Theorem \ref{thm_arnaud1}
          applied with $a=0$:
          \begin{equation*}
            \mus_{p^2}(n) \leq 2n+g_{k+1}-1= 2n+g_k-1 +\Delta g_{k}.
          \end{equation*}
        \item use the algorithm at the step $H_k$ with an appropriate number of
          derivative evaluations. Let $a:= 2(n-n_0^k)$ and suppose that $a \leq
          N_k$. Then\linebreak[4] ${N_k \geq 2n_0^k+2g_k-1}$ implies that ${N_k +a \geq
          2n+2g_k-1}$ so Condition (3) of Theorem \ref{thm_arnaud1} is
          satisfied. Thus, we can perform $a$ derivative evaluations in the
          algorithm using the step $H_k$ and we have:
          \begin{equation*}
            \mus_{p^2}(n) \leq 2n+g_k-1+a.
          \end{equation*}
      \end{enumerate}
      Thus, if $a \leq N_k$, then case (b) gives a better bound as soon as
      ${a<\Delta g_k}$.
      
      For ${x \in \mathbb{R}^{+}}$ such that ${N_{k+1} \geq 2[x]+2g_{k+1}-1}$
      and ${N_k < 2[x]+2g_k-1}$, we define the function $\Phi_k(x)$ as follows:
      \begin{equation*}
        \Phi_k(x) = \left\{\begin{array}{ll}
        2x+g_k-1+2(x-n_0^k) & \mbox{if } 2(x- n_0^k) < \Delta g_k\\
        2x+g_{k+1}-1& \mbox{else}.
      \end{array} \right.
      \end{equation*}
      Note that when case (b) gives a better bound, that is to say
      when\linebreak[4]${2(x-n_0^k) < \Delta g_k}$, then according to Lemma
      \ref{lemme_deltagsr} we also have that
      \begin{equation*}
        2(x-n_0^k)< N_k
      \end{equation*}
      so that we can proceed as in case (b) since there are enough rational
      places at which we can take $a=2(x-n_0^k)$ derivative evaluations on.

      We define the function $\Phi$ for ${x\geq0}$ to be the minimum of the
      functions $\Phi_k$ for which $x$ is in the domain of $\Phi_k$. This
      function is piecewise linear with two kinds of pieces: those which have
      slope $2$ and those which have slope~$4$. Moreover, since the y-intercept
      of each piece grows with $k$, the graph of the function $\Phi$ lies below
      any straight line that lies above all the points
      ${\big(n_0^k+\frac{\Delta g_k}{2}, \Phi(n_0^k+\frac{\Delta
      g_k}{2})\big)}$, since these are the vertices of the graph. If we let
      ${X:=n_0^k+\frac{\Delta g_k}{2}}$, then
      \begin{eqnarray*}
        \Phi(X) & \leq & 2X + g_{k+1} -1 \leq 2\left(1 +
        \frac{g_{k+1}}{2X}\right)X.
      \end{eqnarray*}
      We want to give a bound for $\Phi(X)$ which is independent of $k$.

      Lemmas \ref{lemme_genregsr} ii), \ref{lemme_deltagsr} and
      \ref{lemme_bornesupgsr} give
      \begin{eqnarray*}
        \frac{g_{k+1}}{2X} & \leq & \frac{2^{k+2}}{2^{k+1}(p-3) + 4 + 2^{k+1} -
          2^\frac{k+1}{2}}\\
        & = & \frac{2^{k+2}}{2^{k+1} \left((p-3) + 1 + 2^{-k+1} -
          2^{-\frac{k+1}{2}}\right)}\\
        & = & \frac{2}{p - 2 + 2^{-k+1} - 2^{-\frac{k+1}{2}}}\\
        & \leq & \frac{2}{p - \frac{33}{16}}
      \end{eqnarray*}
      since $-\frac{1}{16}$ is the minimum of the function ${k \mapsto
      2^{-k+1}-2^{-\frac{k+1}{2}}}$. Thus the graph of the function $\Phi$ lies
      below the line ${y=2\left(1+  \frac{2}{p-\frac{33}{16}}\right)x}$. In
      particular, we obtain
      \begin{equation*}
        \Phi(n) \leq 2\left(1+ \frac{2}{p-\frac{33}{16}}\right)n.
      \end{equation*}
    \end{enumerate}
  \qed
\end{Proofof}

\section{Note on some unproven bounds}\label{sect3}

In this section, we discuss a result in the paper \cite{cacrxiya} that to
us seems to be still unproven, and the consequences of this gap for some
asymptotic bounds that were based on this assertion.

\subsection{The result in question}

The following assertion is a folklore conjecture. It states that there exist
curves which, seen over an extension of the base field, have many points. In
the form \cite[Lemma IV.4]{cacrxiya}, it is given as follows:

\begin{ass} \label{assertion}
  Let $p$ be a prime number. For each even positive integer $2 t$, there exists
  a family $X_s$ of curves:
\begin{enumerate}[(i)]
  \item defined over $\F_p$;
  \item whose genera tend to infinity and grow slowly: $g_{s+1} / g_s
    \longrightarrow 1$;
  \item whose number of $\F_{p^{2t}}$-points is asymptotically optimal (i.e.
    the ratio of this number to the genus tends to $p^t-1$).
\end{enumerate}
\end{ass}
Thus, by Lemma~IV.3 of the same paper \cite{cacrxiya}, the family $X_s$ attains
the generalized Drinfeld-Vladut bound for the number of points of degree $2t$.
The paper \cite{cacrxiya} claims to give a proof for this assertion.

\subsection{Our criticism in a nutshell}

The main problem in reading of \cite[Lemma IV.4]{cacrxiya} is that the claims
in the proof of said Lemma are not only highly ambiguous, but also incorrect in
general. The Shimura curves considered in \emph{loc. cit.} have Atkin-Lehner
automorphisms, which in general leads to descent obstructions and the existence
of twists. These issues are not dealt with or even mentioned by the authors,
who state without proof that their Shimura curves are defined over $\Q$. This
forms a sufficiently serious problem to invalidate their proof, at least in our
analysis so far.

In what follows, we will discuss how we have tried to read the claims by the
authors in the most canonical way possible, which leads to the following claim:

\begin{itemize}
  \item[\textbf{Claim A.}] The canonical model of a Shimura curve descends to
    $\Q$.
\end{itemize}
In general, Claim A is incorrect; we give a counterexample below in Section
\ref{sec:ces}. Note that the results \cite[Theorem IV.6, Theorem IV.7,
Corollary IV.8]{cacrxiya}, \cite[Theorem 5.18, Corollary 5.19]{cacrxing3} and \cite[Theorem 5.3, Corollary 5.4, Corollary 5.5]{pira} depend on the aforementioned Lemma
IV.3 of \cite{cacrxiya}.

\subsection{Hypotheses of the Lemma and some further restrictions}

Here we describe the complex analytic quotients of \cite{cacrxiya}. We will
narrow our hypotheses as we go, even beyond those in \emph{loc. cit.} This is
both in order to simplify the presentation and to exclude some cases in which
the statement of the Lemma is clearly false (such as those in which the quaternion
algebra is ramified at primes over $p$).
\begin{itemize}
\item Again, $p$ is any prime number (the one by which the curve is to be
  reduced), and $t$ any integer ($2t$ being the degree for which one wants
  the reduction to have an optimal number of $\F_{p^{2t}}$-rational points).
\item We fix any totally real field $F$ of degree $t$, in which $p$ is inert.
  Choose an embedding $\iota_\infty : F \hookrightarrow \R$, under which $F$
  will be seen as a subfield of $\R$.
\item Finally, fix any given set of finite places $\mathfrak{p}_i$ of $F$ not
  above $p$,\footnote{Notice that the authors do not exclude discriminants
  $\mathfrak{D}^f$ with support meeting $p$. Furthermore, in the cases the
  parity condition allows this, the authors even suggest to choose the
  discriminant equal to the set of primes above $p$. (Note that $p$ being
  inert, this set has one element). Thus, everything is made for the Shimura
  curves to have bad reduction at $p$ (see for example \cite[Theorem
  3.1.6]{sij13}). But this contradicts what is stated later in the paper. So
  we will not consider this case. What the authors might actually have meant
  here is: $\mathfrak{D}^f$ equal to a prime \emph{not} above $p$ (this is
  exactly the requirement asked by \cite{shtsvl}, to which the authors refer
  for this suggestion).} provided that their number plus $t-1$ is even. Call
  $\mathfrak{D}^f$ their product.
\item Now, consider $B$ the quaternion algebra over $F$ which is ramified at
  exactly every real place other than $\iota_\infty$ and all the finite
  places in $\mathfrak{D}^f$.
\item We impose the following further requirements, the first of which is
  demanded in \cite{cacrxiya} as well:
  \begin{itemize}
   \item $\mathfrak{D}^f$ is Galois-invariant.
   \item  $B$ has one single conjugacy class of maximal orders (a sufficient
     condition for this being that $F$ has narrow class number 1).
  \end{itemize}
  The corollary of \cite{dona} then implies that the Shimura curves
  considered here will have field of moduli equal to $\Q$ (if it were not the
  case, then the curves would certainly not descend to $\Q$).
\item Choose a maximal order $\co$ in $B$. Finally, consider the action of
  the subgroup of norm one units $\co^1$ on the upper half-plane\footnote{To
  make things simple, we do not consider level $l$ suborders of $\co$ here
  (which means we consider only the case $l=1$).}, induced by
  \begin{equation*}
    \co^1 \xrightarrow{\iota_\infty} SL_2(\R) \xrightarrow{\text{mod} \pm}
    PSL_2(\R)
  \end{equation*}
  and call $Y^1_0$ the corresponding compact complex analytic quotient.
\item Finally, counterexamples are even simpler if one restricts to
  fields $F$ with narrow class number 1. Indeed, under this additional
  condition, the curves $Y_0^1$ then coincide with $Y_0^+$ (see
  \cite[Proposition 3.2.1]{sij} and the survey below). Thus, they have
  canonical models available in the literature, and moreover defined over
  $F^\infty=F$.
\end{itemize}

\subsection{Classical results on Shimura curves}

A first line of ideas originates from the main theorem of Shimura \cite[Theorem I
\S 3.2]{shi67}, that gives canonical models for certain quotients of the upper
half plane, $Y^+(l)$, which have good reduction above $p$ (by the main result
of \cite{mor}). These models are defined over the ray class
$\mathrm{Cl}((l).\infty)$-extension of $F$. With the help of canonical
models of Shimura for more general quotients of the upper half plane
\cite{shi70}, Ihara builds a family of curves with arbitrary large genus,
smooth over the ring of $(p)$-adic integers of $F_{(p)}$ \cite[\S 6]{ihara79};
their reduction have an asymptotically optimal number of $\F_{p^{2f}}$-points
(see \cite[(1.4.3)]{ihara79} and also the later note \cite{ihara81}).

The second line of ideas uses the construction of Deligne. It can be
illustrated following \cite{duc} and \cite{sij13}. Fix the following notations:
\begin{itemize}
  \item $\co(l)\subset \co$ the Eichler suborders of level $l$ of the paper
    (maximal at every finite place, except at the inert prime $(l)$, where they
    are upper-triangular modulo $(l)$);
  \item $B^{+}$ (resp. $\co(l)^+$) the quaternions (resp. the elements of the
    order) with totally positive norm;
  \item $\mathbf{A}^f$ the finite adeles and $\cH^{\pm}$ the union of the upper
    and lower half-planes. 
\end{itemize}
Consider the double coset space:  
\begin{equation*}
  Y(\co(l)_{\mathbf{A}^f}^{\point})=B^{\point} \backslash \cH^{\pm}\times
  B_{\mathbf{A}^f}^{\point} / \co(l)_{\mathbf{A}^f}^{\point}
\end{equation*}
on which $B^{\point}$ acts on $\cH^{\pm}$ and $B_{\mathbf{A}^f}^{\point}$ on
the left, and $\co(l)_{\mathbf{A}^f}^{\point}$ acts on
$B_{\mathbf{A}^f}^{\point}$ on the right. This space is compact (see
\cite[Example 3.4]{milw}) and has a familiar decomposition in connected
components. Indeed, consider representatives $b_i$ for the quotient $B^{+}
\backslash B_{\mathbf{A}^f}^{\point} / \co(l)_{\mathbf{A}^f}^{\point}$ Then, we
have (see \cite[Lemma 5.13]{milw}):
\begin{equation*}
  Y(\co(l)_{\mathbf{A}^f}^{\point})\cong \bigcup_i Y(b_i\co(l)^{+}b_i^{-1}),
\end{equation*}
where $Y(b_i\co(l)^{+}b_i^{-1})$ stands for $(b_i\co(l)^{+}b_i^{-1})\backslash
\cH$. One has a canonical model over $F$ for the total (non connected) curve
$Y(\co(l)_{\mathbf{A}^f}^{\point})$ (see \cite[\S 1.2]{cara}); over the narrow
class field, this canonical model becomes a product of conjugates of the
component $Y(\co(l)^{+})$ containing $[i, 1]$, as in  \cite[(2.9)]{sij13}.
(Note that because we are dealing with an Eichler order we indeed have that
$F = F^\infty$, in light of \cite[Theorem 1.2.1]{sij13}.) One concludes by
using the fact that $Y(\co(l)_{\mathbf{A}^f}^{\point})$ has good reduction mod
$p$ with many $\F_{p^{2f}}$-points, by \cite[\S 11.2 Remarque (3)]{cara}.

This is the approach of Zink (who studies the reduction of more general
canonical models by hand). On the contrary, the present paper uses the more
computable-friendly curves $Y(\co(l)^{1})$ (popularized by \cite{elk1}). They
occur as coverings of the $Y(\co(l)^{+})$ but they are actually also
encompassed by the same theory (in a non-canonical manner, see the tweak
described in \cite[\S 3.2]{sij}).

\subsection{The main issue: field of definition versus field of moduli}

The paper \cite{cacrxiya} states without further ado that the Shimura curves
$Y(\co(l)^{+})$ of the previous section are defined over $\Q$. While we do not
have a counterexample to this statement, it seems unlikely to hold true in
general. It may be true that their field of moduli is $\Q$, but since the
curves $Y(\co(l)^{+})$ typically have non-trivial automorphisms (namely
Atkin-Lehner involutions), there is always a risk that a descent
obstruction occurs, and we expect that in general this will happen.

Even if the curve did descend, the resulting models would admit twists, which
is to say that there would exists curves over $\Q$ isomorphic with the chosen
model over $\C$ but not over $\Q$. In particular, the statement in \emph{loc.
cit.} that the model over $\Q$ has good reduction modulo $p$ is meaningless,
since it depends on the choice of model.

Moreover, in order to obtain many points over quadratic extensions, we need the
model over $\Q$ to be related with the canonical model in the sense of Shimura.
Read in this way, \emph{loc. cit.} seems to suggest that the canonical model
admits a descent to $\Q$. Thus we end up with Claim A above. While it would
solve the problem, the statement of that claim is false in general, as we now
proceed to show.

\subsection{Counterexamples to descent of the canonical model}\label{sec:ces}

The following table summarizes the properties of three such counterexamples.
The left-hand column is a reference for the data for each of the three curves,
as given in the tables of \cite{sij13}. The second and last columns give the
number field $F$ and the discriminant $\mathfrak D^f$ defining the quaternion
algebra as above (where, for example, $\mathfrak p_3$ and $\mathfrak p_3'$
stand for the two primes over the split prime 3). The two columns in the middle
describe whether the primes $2$ and $3$ are inert in $F$.\footnote{Thus, one
can see that we are unlucky because these counterexamples would not have been,
anyway, good candidates for reduction modulo or 3 (these primes either meet the
discriminant, or they are not inert).}
\enlargethispage{2em}
\begin{table}[!htbp] \caption{Counterexamples} \label{table_cex}
  \begin{center}
    \begin{tabular}{|c|c|c|c|c|} \hline
    curve & F & 2 inert & 3 inert & $\mathfrak{D}^f$ \\  \hline
    e2d13D4 & \multirow{2}{*}{$\Q(\sqrt{13})$} & \multirow{2}{*}{yes} &
      \multirow{2}{*}{no} & $\p_2$ \\ \cline{1-1} \cline{5-5}
    e2d13D36 &  &  &  &  $\p_2\p_3\p_3'$\\ \hline
    e3d8D9 & $\Q(\sqrt{2})$ & no & yes & $\p_3$ \\ \hline
    \end{tabular}
  \end{center}
\end{table}

For these three curves of genus 1, the canonical models, defined over $F$, do
not descend to $\Q$.

\subsection{Proof for one counterexample}
Let $X$ be the curve e2d13D36. $X$ is defined over $F$, of genus 1, but doesn't
necessarily have a rational point. However, we were able to derive properties
of its jacobian $J$, which is an elliptic curve over $F$:
\begin{itemize}
  \item Its conductor equals $6$, by \cite[Proposition 2.1.6]{sij13}.
  \item The valuation of its $j$-invariant at $\mathfrak p_2$ is equal to -10
    (resp. -2 at $\mathfrak p_3$ and $\mathfrak p_3'$). Let us detail this
    result for the valuation at $\mathfrak p_2$. First, define the quaternion
    algebra $H$ ramified exactly at both infinite places of $F$ and at
    $\mathfrak p_3 \mathfrak p_3'$. Call $\co_H$ the maximal order of $H$. As
    in \cite[Proposition 3.1.9 (ii)]{sij13} , consider $\co_H(\mathfrak p_2)$,
    a level $\mathfrak p_2$ suborder of $\co_H$. Consider the set of classes of
    right ideals of $\co_H(\mathfrak p_2)$, noted
    $\mathrm{Pic}_r(\co_H(\mathfrak p_2))$. To each ideal class
    $[\mathrm{I}(\mathfrak p_2)]$ in this set, associate the weight
    $|\co_l(\mathrm{I}(\mathfrak p_2))^{\point}/\Z_{F}^{\point}|$ (equal to the
    cardinality of the projectivized group of units of the left-order of
    $\mathrm{I}(\mathfrak p_2)$). These weights can be computed by running the
    Magma (\cite{magma}) file PadInit in \cite{sijalgo}. The sum of these
    weights is then equal to the opposite of the valuation of $j$ at $\mathfrak
    p_2$, by \cite[Proposition 3.1.14 (iii)]{sij13}. 
\end{itemize}

Now if the curve $X$ were defined over $\Q$, then the jacobian $J$ would
descend to an \emph{elliptic curve} $J_\Q$ over $\Q$, by the argument of
\cite[Proposition 1.9]{mil86}. So, let us suppose that such a rational model
$J_\Q$ does exist, then
\begin{itemize}
  \item the conductor of $J_\Q$ is either equal to $6$, or to $6\cdot13^2$.
    Indeed:
  \begin{itemize}
    \item at every place $p$ but $13$, the extension $F_{\mathfrak P}\Q_p$ does
      not ramify, so the conductor of $J_\Q$ has the same valuation than that
      of $J$, by Proposition 5.4 (a) of \cite{sil86}. (As regards the
      particular cases of $2$ and $3$, note that $J$ has multiplicative
      reduction at these places, so the valuation of the conductor of $J_\Q$ is
      necessarily equal to 1 at these places.)
    \item at the place $13$ where the extension $F_{\mathfrak P}/\Q_{13}$
      ramifies, $J_{\Q}$ cannot have multiplicative reduction. For that if it
      were the case, then $J$ would also have multiplicative reduction at $13$
      (by \cite[Proposition 5.4 (b)]{sil86}). This contradicts the above result
      on the conductor of $J$.
  \end{itemize}
  \item the $j$-invariant of $J_\Q$ should be equal to the one of $J$. So, in
particular, it should have the valuations at $2$ and $3$ predicted above.
\end{itemize}

Then, by a lookup in the tables of Cremona (proved to be exhaustive, see the
introduction of \cite{cre}), only two elliptic curves $E_1$ and $E_2$ over $\Q$
fulfill the conditions above:
\begin{align*}
  y^2 + xy + y & = x^3 - 70997x + 7275296 \\
  y^2 + xy & = x^3 - 11998412x + 15995824272
\end{align*}

But considered over $F$, neither of their conductors is equal to $6$ (one
obtains isomorphic curves over $F$ of conductor $6.13$). So neither of them can
be $J_\Q$, which therefore does not exist.

\subsection{Alternative verifications}
In \cite[Chapter 7]{sij}, the fourth author showed that the canonical model of
$J$ over $F$ is given by
\begin{equation}\label{eq:canmod}
  y^2 + (r + 1) x y + (r + 1) y = x^3 + (16383 r - 38230) x + (1551027 r -
  3576436),
\end{equation}
where $r$ is a root of $t^2 - t - 3$. Explicit methods to verify this equation
were also furnished in \cite{sij}. While these already show the correctness of
the equation (\ref{eq:canmod}), we performed some additional sanity checks:

\begin{itemize}
  \item First, we checked that every quadratic twist of this model involving
    ${\mathfrak p}_2$, ${\mathfrak p}_3$ and ${\mathfrak p}'_3$, leads to a
    strict increase of the actual conductor $6$, so cannot be a candidate for
    $J$.
  \item In addition, we compared the traces of Frobenius on $J$ at several
    primes, to those predicted by the isomorphism (5.16) of \cite{sij13}. This
    isomorphism asserts that the representation of the Hecke algebra on the
    (one-dimensional) space of differentials on $E$, is isomorphic to the
    representation of the Hecke algebra on the space of $\mathfrak D^f$-new
    Hilbert cusp forms on $F$. The comparison was made possible, since the
    traces for this last representation are also computable in Magma (by the
    work of Dembélé and Donnelly \cite{dedo}).
\end{itemize}

Now, as remarked above, to show that the curve e2d13D36 is not defined over $\Q$, it suffices to show that the jacobian $J$ does not descend to an \emph{elliptic} curve over $\Q$. The equation for $J$ given above (\ref{eq:canmod}) enables one to check this fact directly. For example, here are two ways to see it : \begin{itemize}
  \item The trace of the Frobenius of $J$ at the inert prime $(11)$, is equal to $22$, which is not of the
form $n^2 - 2 \cdot 11$.
  \item Alternatively, one can check that the Weil cocycle criterion is not
    satisfied for the curve $J$. Namely, denoting the conjugation of the quadratic field $F$
    by $\sigma$, this boils down to verifying that, for any $F$-isomorphism
    $f_\sigma : J \rightarrow J^\sigma$ from $J$ to the conjugate curve,
    then $f_\sigma$ does not satisfy $f_\sigma \circ \sigma(f_\sigma) =
    \mathrm{id}$. The automorphism group of the elliptic curve $J$ being of
    order 2, this is quickly done.
\end{itemize}

 Finally, there exists a last -- and more straightforward -- way to prove that
 e2d13D36 is a counterexample. It does not use the actual equation for the
 canonical model $J$, nor appeals to the various sophisticated theories used
 above (that predict the traces, conductor and $j$-invariant). This approach
 consists in computing the traces of the Hecke operators on $J$ in the direct
 manner. Namely, \cite[Algorithm 4.2.1]{sij} (available in \cite{sijalgo},
 TakData) enables one to compute the action of the Hecke operators on the
 homology of the complex curve $Y_0^1$. Then, the computation of the trace at
 the inert prime $(11)$ leads to the same result, and thus conclusion, as
 above.

\subsection{Further exegesis}

This concludes our discussion of the claims made in \cite[Lemma
IV.4]{cacrxiya}. We consider the proof of that Lemma as essentially flawed.
That said, it seems likely that there are yet ways in which the result can be
salvaged, which requires a finer analysis of the automorphism groups and the
cohomological descent problems encountered. We hope to deal with these issues
in the future.

\section*{Acknowledgment}

The authors wish to thank the anonymous referee for his constructive and
valuable comments which helped a lot to improve the manuscript.

\bibliographystyle{amsplain}

\end{document}